\documentclass[aps, prb, twocolumn, 10pt, superscriptaddress, floatfix]{revtex4-2}
\usepackage{amsmath}
\usepackage{amssymb}
\usepackage{braket}
\usepackage{mathtools}
\usepackage{graphicx}
\usepackage{color}
\usepackage{xcolor}
\usepackage{ulem}
\usepackage{xspace}
\usepackage{upquote}
\usepackage{csquotes}
\usepackage{tabularx,makecell}
\usepackage{bm,dsfont}
\usepackage[utf8]{inputenc}
\usepackage[vietnamese, english]{babel}
\usepackage{chngcntr}
\usepackage{xcolor}
\usepackage{braket}
\usepackage{physics}
\usepackage{sidecap}
\usepackage{lipsum}
\usepackage{amsfonts}
\usepackage{bbold}
 \usepackage{tensor}
\usepackage[utf8]{inputenc}
\usepackage[vietnamese, english]{babel}

\DeclareUnicodeCharacter{0650}{\unskip}

\usepackage[bookmarks=true,colorlinks=true,urlcolor=blue,linkcolor=blue,citecolor=blue,breaklinks]{hyperref}
\usepackage{booktabs}

\graphicspath{{Fig/}}

\begin{document}

\title{ِSqueezing Quantum States in  Three-Dimensional Twisted Crystals}

\author{\foreignlanguage{vietnamese}{Võ Tiến Phong}}
\affiliation{Department of Physics, Florida State University, Tallahassee, FL, 32306, U.S.A.}
 \affiliation{National High Magnetic Field Laboratory, Tallahassee, FL, 32310, U.S.A.}
\author{Kason Kunkelmann}
\affiliation{Department of Physics and Astronomy, University of Pennsylvania, Philadelphia, Pennsylvania 19104, U.S.A.}
\author{Christophe De Beule}
\affiliation{Department of Physics and Astronomy, University of Pennsylvania, Philadelphia, Pennsylvania 19104, U.S.A.}
\author{Mohammed M. Al Ezzi}
\affiliation{Department of Materials Science and Engineering,
National University of Singapore, 9 Engineering Drive 1,
Singapore 117575}
\affiliation{Centre for Advanced 2D Materials, National University of Singapore, 6 Science Drive 2, Singapore 117546}
\affiliation{Department of Physics, Faculty of Science, National University of Singapore, 2 Science Drive 3, Singapore 117542}
\author{Robert-Jan Slager}
\affiliation{TCM Group, Cavendish Laboratory, J.J. Thomson Avenue, Cambridge CB3 0HE, U.K.}
\author{Shaffique Adam}
\affiliation{Department of Physics, Washington University in St. Louis, St. Louis, Missouri 63130, United States}
\affiliation{Department of Physics and Astronomy, University of Pennsylvania, Philadelphia, Pennsylvania 19104, U.S.A.}
\affiliation{Department of Materials Science and Engineering,
National University of Singapore, 9 Engineering Drive 1,
Singapore 117575}
\author{E. J. Mele}
\affiliation{Department of Physics and Astronomy, University of Pennsylvania, Philadelphia, Pennsylvania 19104, U.S.A.}

\date{\today} 

\begin{abstract}
\normalsize
Bloch's theorem provides a conventional starting point for describing wave propagation in  periodic media but in ordered materials where competing spatial periods coexist it is rendered ineffective, often with dramatic consequences. Here we develop an alternate approach that uses coherent free particle vortex states to study quantum states in supertwisted crystals: three dimensional stacks of atomically thin two dimensional layers. This formalism leads naturally to the representation of the spectrum using squeezed coherent states and reveals the crucial role of a Coriolis coupling in the equations of motion. This identifies an underlying noncommutative geometry and novel edge state structure in a family of complex ordered structures.

\end{abstract}

\maketitle

\section{Introduction}
A fundamental idea in wave mechanics is that propagation in a periodic medium can be described using Bloch's theorem: propagating waves are  indexed by their conserved crystal momenta  that label their transformations when displaced by a set of discrete lattice translations.  In ordered materials where incommensurate spatial periods compete, this general principle is often rendered ineffective. Examples are crystals with broken symmetries from charge or spin density waves \cite{monceau_electronic_2012,takayama_continuum_1980}, quasiperiodic lattices that produce diffraction patterns with crystallographically forbidden point symmetries \cite{yu_dodecagonal_2019,rotenberg_electronic_2004}, and stacks of two-dimensional lattices with a relative rotation (twist) between  layers \cite{andrei_marvels_2021}. In special cases when there is a small difference between the competing periods, a  useful work-around has been to adopt a continuum description where a periodic long-wavelength field produces Bragg scattering that coherently mixes  short-wavelength carrier waves   \cite{brazovskii_exact_1980,bistritzer_moire_2011}. In this work, we advocate an alternative approach to study three-dimensional twisted crystals that replaces their spectrally congested momentum-space Bloch band structures \cite{cea_twists_2019,wu_three-dimensional_2020} with a  representation  using squeezed  states in a Fock space of free-particle vortex states \cite{bliokh_theory_2017}. This reorganization of the Hilbert space highlights the role of the Coriolis term in the equations of motion that produces unconventional phase space dynamics and edge state structure generic to a family of complex crystals.

Screw symmetry of a twisted crystal allows one to separate modes into invariant sectors labeled by a screw eigenvalue $\kappa_z$ \cite{wu_three-dimensional_2020} effectively replacing the plane-wave momentum $k_z$. Importantly, this choice  couples the in-plane and out-of-plane motions and necessitates abandoning the conventional Bloch wave representation of dynamics perpendicular to the screw axis. Instead, each $\kappa_z$ sector inherits a Coriolis coupling to the twist, similar to the 2D motion of a charged particle in a uniform magnetic field, along with an (outward) centrifugal potential. This latter feature distinguishes this system from a related problem that occurs in rotating trapped ultracold atomic gases where the centrifugal potential can be nearly balanced by a confining potential allowing a faithful mapping to Landau level dynamics from the Coriolis coupling \cite{fletcher_geometric_2021,crepel_geometric_2023,mukherjee_crystallization_2022}. By contrast, in a twisted crystal the centrifugal potential plays the crucial role of restoring nearly ``flat" two-dimensional dynamics to a system that would otherwise be  spatially confined by the Coriolis deflection. We find that augmenting the twisted free particle Hamiltonian to include a  crystal potential leads to two classes of quantum states: (i) a spectrally isolated low energy sector with extended states that propagate through saddle points in the crystal potential and (ii) a high energy dispersive sector where states ride over the potential landscape and propagate freely. At lateral boundaries a spectral gap between these  two sectors is traversed by helical edge states in which reversed orbital angular momentum states counterpropagate ballistically in a pattern of winding edge channels. In the coherent state representation presented below all these features are identified as the generic signatures of wave motion in  screw symmetric media. 

\section{Vortex State Representation}
\label{sec: Landau-Level Representation of the Free-Particle Hamiltonian}

In this section we project a free particle Hamiltonian into invariant sectors identified by  their transformations under a screw symmetry operation. In the following $\beta = \partial \phi/\partial z$ labels the in plane rotation $\phi$ of a symmetry axis of the system as a function of height $z$ i.e. it gives the  ``pitch" of the screw.

In three dimensions, aligning the $z$ direction with the screw axis, the free particle kinetic energy can be projected into independent screw sectors $\kappa_z$ with the replacement $\hat p_z = \hbar\kappa_z - \beta \hat L_z,$ where $\hat L_z = \hat x \hat p_y - \hat y \hat p_x$ is the orbital angular momentum operator and $\kappa_z$ is the screw eigenvalue. Completing the squares on the $x$ and $y$ components of the  momenta, we obtain
\begin{equation}
\begin{split}
\label{eq: kinetic energy}
\hat {\mathcal{K}} &=\hat {\mathcal{K}}_\| + \hat V_{\rm c} + \hat {\mathcal{K}}_z, \\
\hat{\mathcal{K}}_\parallel &= \frac{1}{2m} \left[ \left(\hat p_x + \hbar \kappa_z \beta  \hat y\right)^2 + \left(\hat p_y - \hbar \kappa_z \beta  \hat x \right)^2 \right], \\
 \hat{V}_c&= -\frac{\hbar^2 \kappa_z^2 \beta^2}{2m}  \left(\hat x^2 + \hat y^2 \right), \\
 \hat{\mathcal{K}}_z &= \frac{\hbar^2\kappa_z^2}{2m} + \frac{\beta^2\hat{L}_z^2}{2m}.
\end{split}
\end{equation}
Here, $\hat {\mathcal{K}}_\|$ contains a $\kappa_z$-dependent Coriolis coupling to the twist, analogous  to the orbital coupling of a charge $e$ particle to a uniform magnetic field $B^* = 2 \hbar \beta \kappa_z/e$, an effective cyclotron frequency $\omega_c=2 \hbar |\beta \kappa_z|/m$, and a magnetic length $\ell_\beta = 1/\sqrt{2 |\beta\kappa_z |}$. Physically, as a particle moves vertically it encounters a rotating crystal frame which can induce a Coriolis deflection. Since $B^*$ depends on the {\it product} $\beta \kappa_z$, sectors at $\pm \kappa_z$ see reversed effective $B^*$'s.  $\hat{V}_c$ is a $\kappa_z$-dependent, isotropic centrifugal potential that forces particles away from the twist axis. $\hat{\mathcal{K}}_z$ contains the orbital angular momentum along the screw axis.

For a two-dimensional system in a uniform perpendicular magnetic field, the last two terms in Eq. \eqref{eq: kinetic energy} are absent, and, after separation of variables in mutually commuting left-handed (${\hat{\boldsymbol{\pi}}}$) and right-handed (${\hat{\boldsymbol{\kappa}}}$) degrees of freedom, the first term reverts to a Hamiltonian for a two-dimensional harmonic oscillator, as we shall show.  The oscillator eigenstates form a ladder of discrete states (the Landau levels) that inherit extensive orbital degeneracies from the independent (guiding center) degree of freedom.  In rotating systems the situation is more complicated because the centrifugal potential $V_c$ is inevitably present. However,  experiments on rotating quantum atomic gases manage to null  $V_c$  by applying  a trapping potential in order to access the Landau level dynamics \cite{fletcher_geometric_2021,crepel_geometric_2023}.  This simplification does not occur for the analogous problem in a twisted crystal.

We now make the connection to quantum Hall physics explicit by introducing two sets of mutually commuting left-handed $\hat{\boldsymbol{\pi}}$ and right-handed $\hat{\boldsymbol{\kappa}}$ operators:
\begin{equation}
\begin{split}
    \hat{\pi}_x &= \hat{p}_x - \hbar \beta\kappa_z\hat{y} \quad \text{and} \quad    \hat{\pi}_y = \hat{p}_y + \hbar \beta\kappa_z\hat{x}, \\
    \hat{\kappa}_x &= \hat{p}_x + \hbar \beta\kappa_z\hat{y} \quad \text{and} \quad    \hat{\kappa}_y = \hat{p}_y - \hbar \beta\kappa_z\hat{x}. 
\end{split}
\end{equation}
The $\hat{\boldsymbol{\pi}}$ operators are transformed into the $\hat{\boldsymbol{\kappa}}$ operators under exchanging $\beta \kappa_z \mapsto - \beta \kappa_z.$  Consequently, these operators degenerate when $\beta \kappa_z = 0,$ in which case the $\hat{x}$ and $\hat{y}$ operators completely disappear from these expressions. (The $\beta\kappa_z = 0$ case is handled  separately in Appendix \ref{sec: kappaz = 0}). For now, assuming $\beta\kappa_z \neq 0,$ these operators satisfy the following commutation relations:
\begin{equation}
\begin{split}
    \left[ \hat{\pi}_x, \hat{\pi}_y \right]   = -\left[ \hat{\kappa}_x, \hat{\kappa}_y \right] = -  \frac{is \hbar^2}{\ell_{\beta}^2}, \quad    \left[ \hat{\pi}_i, \hat{\kappa}_j \right] = 0.
\end{split}
\end{equation}
We  note that the commutation relations change sign based on $s = \text{sgn}\left[\beta \kappa_z \right].$  We are now ready to define oscillator ladder operators:
\begin{equation}
    \begin{split}
        \hat{a} &= \frac{\ell_\beta}{\sqrt{2}\hbar} \left( \hat{\pi}_x - i s \hat{\pi}_y \right) \quad \text{and} \quad   \hat{a}^\dagger = \frac{\ell_\beta}{\sqrt{2}\hbar} \left( \hat{\pi}_x + i s \hat{\pi}_y \right), \\
        \hat{b} &= \frac{\ell_\beta}{\sqrt{2}\hbar} \left( \hat{\kappa}_x + i s \hat{\kappa}_y \right) \quad \text{and} \quad    \hat{b}^\dagger = \frac{\ell_\beta}{\sqrt{2}\hbar} \left( \hat{\kappa}_x - i s \hat{\kappa}_y \right). 
    \end{split}
\end{equation}
We emphasize that these $\hat{a}$ and $\hat{b}$ operators are defined in each $\kappa_z$ sector (i.e. each $\kappa_z$ gives a different combination of momentum and position operators). If one insists, one can emphasize this dependence by including subscripts: $\hat{a}_{\kappa_z}$ and $\hat{b}_{\kappa_z}$; we leave them off to avoid cluttered notion.  The factors of $s$ have been inserted specifically to ensure these combinations obey the commutation relations:
\begin{equation}
    \begin{split}
        \left[ \hat{a}, \hat{a}^\dagger \right] &= 
        \left[ \hat{b}, \hat{b}^\dagger \right]  = 1,        \\
        \left[ \hat{a}, \hat{b}^\dagger \right] &= \left[ \hat{a}, \hat{b} \right] = \left[ \hat{a}^\dagger, \hat{b}^\dagger \right] = \left[ \hat{a}^\dagger, \hat{b} \right]= 0.
    \end{split}
\end{equation}
Our task now is to write the Hamiltonian in terms of these ladder operators. To do that, let us rewrite the momentum and position operators in terms of the $\hat{a}$ and $\hat{b}$ operators
\begin{equation}
\begin{split}
 \label{eq: equivalence between a b and x y}
    \begin{pmatrix}
        \hat{p}_x \\
        \hbar \beta \kappa_z\hat{x}\\
        \hat{p}_y \\
        \hbar \beta \kappa_z\hat{y}
    \end{pmatrix} &= \frac{\hbar}{2\sqrt{2}\ell_\beta}\left(
\begin{array}{cccc}
 1 & 1 & 1 & 1 \\
 i s & -i s & is & -is \\
 i s & -i s & -is & is \\
 -1 & -1 & 1 & 1 \\
\end{array}
\right)\begin{pmatrix}
        \hat{a} \\
        \hat{a}^\dagger \\
        \hat{b} \\
        \hat{b}^\dagger
    \end{pmatrix}.   
\end{split}
\end{equation}
The angular momentum operator is the signed ($s$) difference of the number operators in the $a$ and $b$ sectors
\begin{equation}
    \hat{L}_z = \hbar \hat{M} = s\hbar \left( \hat{a}^\dagger \hat{a} - \hat{b}^\dagger \hat{b} \right).
\end{equation}
The harmonic potential in the ladder representation is 
\begin{equation}
\label{eq: contrifugal potential operator}
    \hat{V}_c = \frac{\hbar\omega_c}{4} \left[ \hat{a} \hat{b} + \hat{b}^\dagger \hat{a}^\dagger - \hat{a}^\dagger \hat{a} - \hat{b}^\dagger \hat{b}-1  \right].
\end{equation}
The in-plane Hamiltonian takes the form of a harmonic oscillator Hamiltonian as mentioned before
\begin{equation}
    \hat{\mathcal{K}}_\parallel = \hbar \omega_c \left(\hat{b}^\dagger \hat{b} + \frac{1}{2} \right).
\end{equation} Using these results, the full Hamiltonian can be written simply as
\begin{equation}
\label{eq: nice form of equation}\hat{\mathcal{K}} = \frac{\hbar \omega_c}{4} \left(     \hat{N} + \hat{a}\hat{b} + \hat{b}^\dagger \hat{a}^\dagger  + 1 \right) +  \frac{\hbar^2}{2m} \left(\kappa_z - \beta \hat{M} \right)^2,
\end{equation}
where $\hat{N} = \hat{n}_a+ \hat{n}_b = \hat{a}^\dagger \hat{a} +  \hat{b}^\dagger \hat{b}$ is the total mode operator. It is worth pointing out that since $\left[ \hat{N}, \hat{M} \right] = 0$ and $\left[\hat{a}\hat{b} + \hat{b}^\dagger \hat{a}^\dagger, \hat{M} \right] = 0,$ the whole Hamiltonian satisfies $\left[ \hat{\mathcal{K}}, \hat{M}\right] = 0.$ Therefore, for the isotropic problem, the eigenstates of $\hat{\mathcal{K}}$ can be labeled by a conserved orbital angular momentum.

\begin{figure*}[ht!]
\centering
\includegraphics[width=\textwidth]{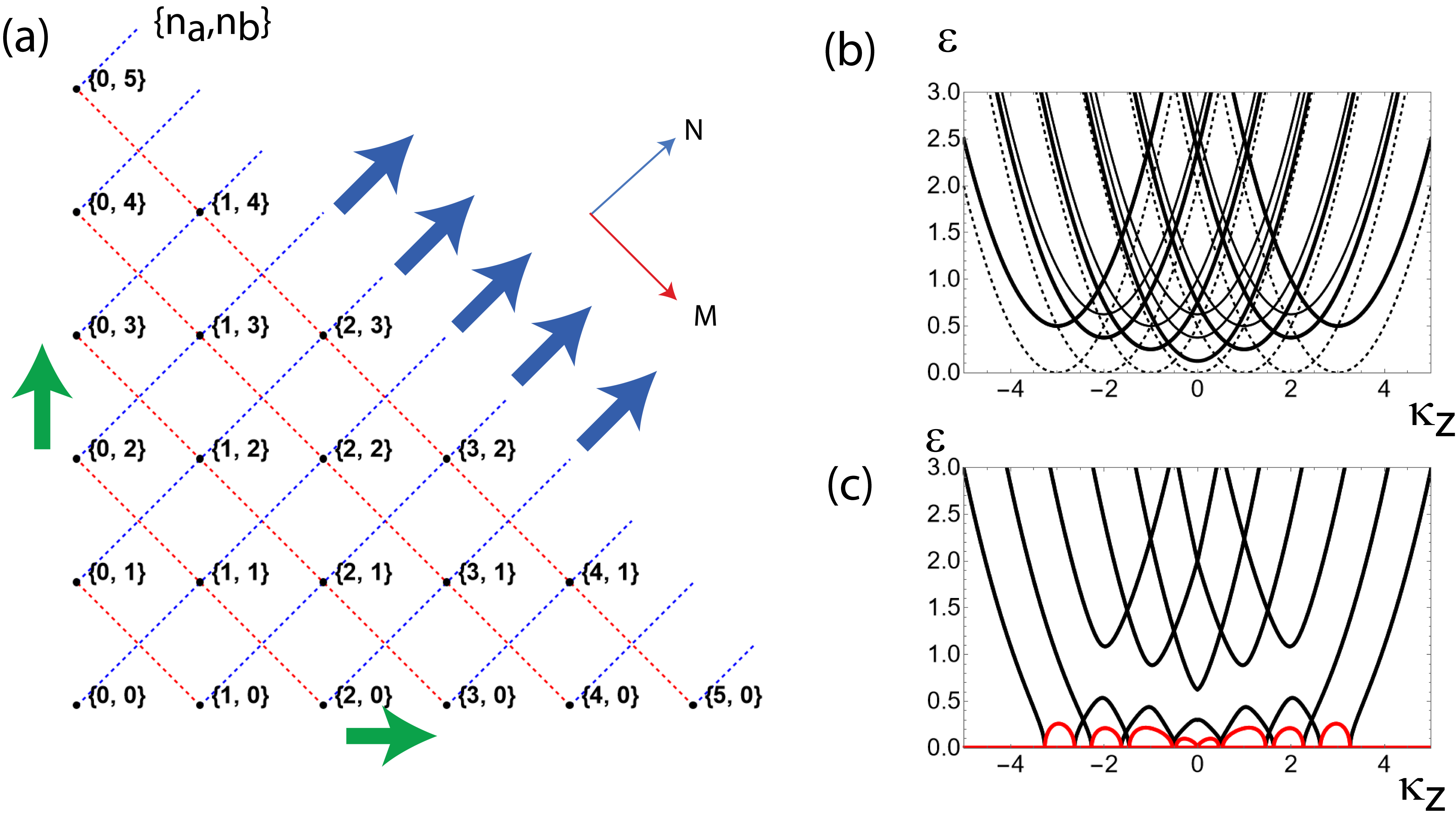}
\caption{ (a) Fock space for quantum states in a three-dimensional twisted crystal labeled by occupation numbers in time reversed angular momentum states $\{n_a, \, n_b \}$. $M = n_a - n_b$ labels the orbital angular momentum and $N = n_a + n_b$ labels the total occupation number.  Green arrows denote sectors that are mixed in a constant magnetic field in the extreme quantum limit. Blue arrows denote the mixing of orbitals in a twisted crystal producing superpositions with different occupation numbers $N$ in strings that conserve angular momentum $M$. (b) (Dashed line) Free electron kinetic energy $\varepsilon = 2mE/\hbar^2 \beta^2$ for motion along the screw axis as a function of the screw eigenvalue (with $\beta = 1$). (Solid) Dispersion of the Bogoliubov eigenstates for the free particle spectrum regularized by introducing a small harmonic potential. (Bold) Minimal states with $N=|M|$ in each angular momentum channel. (c) Gapping of the free particle spectrum by a saddle point potential that mixes modes with $M$ and $M\pm 2$ (here $|M|\le 3$). Overtones of the pinning potential in the complete excitation spectrum have been omitted for clarity (Appendix \ref{sec: Effect of Saddle-Point Potential Energy} discusses this subtlety).
\label{fig: band structure}}
\end{figure*}

\section{Uniformly Twisted Crystal}

The derivation in Sec. \ref{sec: Landau-Level Representation of the Free-Particle Hamiltonian} shows that the free-particle Hamiltonian in invariant sectors with definite screw eigenvalues has an intrinsic connection to Landau-level physics encoded in $\hat{\mathcal{K}}_\parallel.$ Furthermore, the derivation also makes clear that there are additional features unique to this setup encoded in $\hat{V}_c$ and $\hat{\mathcal{K}}_z.$ Therefore, application of this formalism to a  crystal with a uniform twist (constant $\beta$) introduces three new essential considerations. (i) Both ``right" and ``left"-handed sectors become accessible in the Hilbert space. A map of the Fock space of number states $\{n_a,n_b\} = \{\langle \hat a^\dag \hat a \rangle, \langle \hat b^\dag \hat b \rangle \}$ for the two oscillator quanta is shown in Fig. \ref{fig: band structure}(a).   (ii) Propagation along the screw direction disperses the bands in energy, violating the Landau level flat band condition.  (iii) The centrifugal potential is present and plays the crucial role of opposing the tendency of the oscillator states to localize around a  single (gauge dependent) axis of rotation. We examine these three features in turn.

(i) The number states displayed in Fig. \ref{fig: band structure} are free particle Laguerre-Gauss vortex states \cite{bliokh_theory_2017} which can be written (using complex lateral coordinates ${\mathfrak{z}} \equiv x+iy$ and ignoring $s$ for the moment) as
\begin{equation}
\begin{split}
    \psi_{n_a,n_b}^{LG}  &=  A e^{{\mathfrak{z}} \bar {\mathfrak{z}}/4 \ell_\beta^2} \left(\frac{\partial}{\partial \bar {\mathfrak{z}}} \right)^{n_a} \left(\frac{\partial}{\partial{\mathfrak{z}}} \right)^{n_b}  e^{-{\mathfrak{z}} \bar {\mathfrak{z}}/2 \ell_\beta^2} e^{i (\kappa_z - M \beta) z}  \\
    &\sim  \, r^{N}  e^{i M \phi} e^{-r^2/4 \ell_\beta^2} e^{i (\kappa_z - M \beta) z} \nonumber
\end{split}
\end{equation}
and provide a complete basis where the  amplitude for a state with angular momentum $M=n_a-n_b$ and total number $N=n_a+n_b$ is peaked on spiral paths with mean projected radius $\bar r_N = \sqrt{2N} \ell_\beta$.  These states are related to superposition states produced in free electron vortex beams \cite{bliokh_theory_2017} and in light beams with orbital angular momentum \cite{allen_orbital_1992}. They are ``ultra-localized" around a putative axis of rotation, being confined by the gauge field and have a Gaussian decay in the far field determined by the $\kappa_z$-dependent magnetic length.

(ii) The energy dispersion associated with motion along the stacking direction for a state with definite $M$ is given by (completing the square for the last two terms in Eq. \eqref{eq: kinetic energy}) $
     (\hbar^2/2m)  \left(\kappa_z - \beta M \right)^2 $
which provides a $\kappa_z$-dependent effective potential for a  lateral two-dimensional projected problem (dashed curves in Fig. \ref{fig: band structure}(b)). This separates the free particle spectrum into overlapping branches with zero energy states at shifted twist momenta $\kappa_z = \beta M$. At these special values, the phases accumulated by the in-plane and out of plane motions exactly cancel giving  states with minimal kinetic energy.

(iii) The centrifugal potential has a bilinear form in  bosonic raising and lowering operators (from Eqs. \ref{eq: contrifugal potential operator} and \ref{eq: nice form of equation}, repeated here for clarity)
\begin{equation}
    - \frac{\hbar^2 \kappa_z^2 \beta^2}{2m} \left(\hat{x}^2 + \hat{y}^2 \right) = \frac{\hbar \omega_c}{4} \left(\hat a^\dag \hat b^\dag + \hat a \hat b- \hat a^\dag \hat a - \hat b^\dag \hat b - 1 \right)  \nonumber
\end{equation}
which provides a compact expression for the kinetic energy operator in Eq. \eqref{eq: kinetic energy}
\begin{equation}
\label{eq: kinetic energy in operator form}
\hat {\mathcal{K}} = \frac{\hbar \omega_c}{4} \left[\hat a^\dag \hat a + \hat b^\dag \hat b + 1 + \hat a^\dag \hat b^\dag + \hat a \hat b \right] + \frac{\hbar^2}{2m}  \left[\kappa_z - \beta M \right]^2 
\end{equation}
The first term  is a quadratic form $
    \hat{\mathcal{K}}_\text{quad} = (\hbar \omega_c/4)  (\hat a^\dag + \hat b)(\hat a + \hat b^\dag) = (1/2m) \, \hat p_+ \hat p_-$ 
verifying that Eq. \eqref{eq: kinetic energy in operator form} recovers the free particle kinetic energy. In the absence of scattering from a twisted potential (whose effects will be examined below) free particles  travel in straight lines.  Note also that Eq. \eqref{eq: kinetic energy in operator form} restores the symmetry in the first term between the ladder operators for the $a$- and $b$- quanta that is necessarily missing from the pure magnetic field problem.  This symmetry is ultimately broken by the coupling of $M$ to $\kappa_z$ in the second term demonstrating that a Coriolis deflection requires a physical potential that produces a screw axis $\beta \neq 0$ seen in a sector with $\kappa_z \neq 0$ through the cross term  $-2 \kappa_z \beta M$.

The Hamiltonian in \eqref{eq: kinetic energy in operator form} is bilinear in the bosonic operators and can be brought to diagonal form by a Bogoliubov transformation \cite{colpa1978diagonalization, van1980note, Gurarie2003Bosonic, Complex2005Rossignoli}.  Reordering  the operators using the commutation relations, the quadratic part of the kinetic energy can be rewritten as follows:
\begin{equation}
\label{eq: reordered operators}
\begin{split}
    \hat{\mathcal{K}}_\text{quad} 
&= \frac{\hbar\omega_c}{8} \begin{pmatrix}
    \hat{a}^\dagger & \hat{b} & \hat{b}^\dagger & \hat{a}
\end{pmatrix} \begin{pmatrix}
    1 & 1 & 0 & 0 \\
    1 & 1 & 0 & 0 \\
    0 & 0 & 1 & 1 \\
    0 & 0 & 1 & 1
\end{pmatrix}\begin{pmatrix}
    \hat{a} \\ \hat{b}^\dagger \\ \hat{b} \\ \hat{a}^\dagger
\end{pmatrix}.
\end{split}
\end{equation}
We note that this Hamiltonian has a block-diagonal structure because there are no terms of the sort $\hat{a}\hat{a}$ and $\hat{b}\hat{b}$. A bosonic Bogoliubov transformation is defined by the properties of a symplectic group. Williamson’s Theorem states that for any real $2n\times 2n$ positive-definite matrix, one can find a symplectic transformation such that the symplectic eigenvalues are positive-definite \cite{nicacio2021williamson}. Unfortunately, the matrix in Eq. \eqref{eq: reordered operators} is not positive-definite and so, we need to regularize it by replacing the off-diagonal elements by $|\lambda| < 1.$ This regularization turns out to be equivalent to imposing a (weak) confining potential. With this regularization, we implement the following Bogoliubov transformation \cite{Gurarie2003Bosonic}:
\begin{equation}
    \begin{pmatrix}
        \hat{a} \\
        \hat{b}^\dagger \\ 
        \hat{b} \\ 
        \hat{a}^\dagger \\ 
    \end{pmatrix} = \begin{pmatrix}
        u & v & 0 & 0 \\
        v & u & 0 & 0 \\
        0 & 0 & u & v \\
        0 & 0 & v & u \\
    \end{pmatrix}\begin{pmatrix}
        \hat{\mathbb{a}} \\
        \hat{\mathbb{b}}^\dagger \\ 
        \hat{\mathbb{b}} \\ 
        \hat{\mathbb{a}}^\dagger \\ 
    \end{pmatrix},
\end{equation}
where we assume that $u$ and $v$ are real numbers.  The commutation relations on $\hat{a}$ and $\hat{b}$ are satisfied if we choose 
\begin{equation}
    \begin{split}
        \left[ \hat{\mathbb{a}}, \hat{\mathbb{a}}^\dagger \right] = \left[ \hat{\mathbb{b}}, \hat{\mathbb{b}}^\dagger \right] = 1,\quad 
        \left[ \hat{\mathbb{a}}, \hat{\mathbb{b}} \right] = \left[ \hat{\mathbb{a}}, \hat{\mathbb{b}}^\dagger \right] = 0 ,\quad
        u^2 - v^2 = 1.
    \end{split}
\end{equation}
The last condition can be satisfied by choosing $u = \cosh \varphi$ and $v = \sinh \varphi,$ where $\varphi$ is a real parameter. The Hamiltonian matrix can be made diagonal by choosing
\begin{equation}
    \lambda  \left(u^2+v^2\right)+2 u v= 0 \rightarrow \tanh \left( 2 \varphi\right) = -\lambda.
\end{equation}
As long as $|\lambda|<1,$ $\varphi$ is real and so are $\cosh \varphi$ and $\sinh\varphi.$ In this case, the inverse transformations are 
\begin{equation}
    \begin{pmatrix}
        \hat{\mathbb{a}} \\
        \hat{\mathbb{b}}^\dagger 
    \end{pmatrix} = \begin{pmatrix}
        \cosh \varphi & - \sinh \varphi \\
        - \sinh \varphi & \cosh \varphi \\
    \end{pmatrix}    \begin{pmatrix}
        \hat{a} \\
        \hat{b}^\dagger \\ 
    \end{pmatrix},
\end{equation}
This transformation preserves the fact that $\hat{\mathbb{a}}^\dagger$ is the Hermitian adjoint of $\hat{\mathbb{a}}$ and $\hat{\mathbb{b}}^\dagger$ is the Hermitian adjoint of $\hat{\mathbb{b}}$ as desired. The quadratic piece of the Hamiltonian has a diagonal form using the new operators 
\begin{equation}
\hat{\mathcal{K}}_\text{quad}  = \frac{\hbar\omega_c \sqrt{1-\lambda^2}}{4} \left(\hat{\mathbb{a}}^\dagger\hat{\mathbb{a}} +  \hat{\mathbb{b}}^\dagger\hat{\mathbb{b}} + 1 \right) .
\end{equation}
Finally, we need to represent the angular momentum $\hat{a}^\dagger \hat{a}- \hat{b}^\dagger\hat{b}$ terms in the new basis 
\begin{equation}
    \begin{split}
        \hat{a}^\dagger \hat{a} & = u^2 \hat{\mathbb{a}}^\dagger \hat{\mathbb{a}} + uv \left(\hat{\mathbb{a}}^\dagger \hat{\mathbb{b}}^\dagger + \hat{\mathbb{b}}\hat{\mathbb{a}} \right) + v^2  \hat{\mathbb{b}}^\dagger \hat{\mathbb{b}}  + v^2, \\
        \hat{b}^\dagger \hat{b} &=  u^2 \hat{\mathbb{b}}^\dagger \hat{\mathbb{b}} + uv \left(\hat{\mathbb{b}}^\dagger \hat{\mathbb{a}}^\dagger + \hat{\mathbb{a}}\hat{\mathbb{b}} \right) + v^2\hat{\mathbb{a}}^\dagger \hat{\mathbb{a}} +v^2 
    \end{split}
\end{equation}
giving
\begin{equation}
     \hat{a}^\dagger \hat{a}- \hat{b}^\dagger\hat{b}= \hat{\mathbb{a}}^\dagger \hat{\mathbb{a}} - \hat{\mathbb{b}}^\dagger \hat{\mathbb{b}}.
\end{equation}
Therefore, in terms of these Bogoliubov operators, we arrive at the following Hamiltonian 
\begin{equation}
\begin{split}
\hat{\mathcal{K}} &= \frac{\hbar\omega_c \sqrt{1-\lambda^2}}{4} \left(\hat{\mathbb{a}}^\dagger\hat{\mathbb{a}} +  \hat{\mathbb{b}}^\dagger\hat{\mathbb{b}} + 1 \right) \\
&+  \frac{\hbar^2}{2m} \left(\kappa_z - \beta s\left[ \hat{\mathbb{a}}^\dagger \hat{\mathbb{a}} - \hat{\mathbb{b}}^\dagger \hat{\mathbb{b}} \right] \right)^2.    
\end{split}
\end{equation}
We notice that the angular momentum operator now has the same form as before 
\begin{equation}
    \hat{M} = s\left(\hat{a}^\dagger \hat{a} - \hat{b}^\dagger \hat{b}\right) = s  \left(\hat{\mathbb{a}}^\dagger \hat{\mathbb{a}} - \hat{\mathbb{b}}^\dagger \hat{\mathbb{b}}\right).
\end{equation}
The eigenstates and their energies are given by
\begin{equation}
\begin{split}
     \ket{n_\mathbb{a},n_\mathbb{b}} &= \frac{\left[\hat{\mathbb{a}}^\dagger\right]^{n_\mathbb{a}}}{\sqrt{n_\mathbb{a}!}}\frac{\left[\hat{\mathbb{b}}^\dagger\right]^{n_\mathbb{b}}}{\sqrt{n_\mathbb{b}!}} \ket{0,0}, \\
     E_{n_\mathbb{a},n_\mathbb{b}}(\kappa_z) &= \frac{\hbar\omega_c \sqrt{1-\lambda^2}}{4} \left(n_\mathbb{a} +  n_\mathbb{b} + 1 \right) \\
     &+  \frac{\hbar^2}{2m} \left(\kappa_z - \beta s\left[ n_\mathbb{a} - n_\mathbb{b} \right] \right)^2.    
\end{split}
 \end{equation}
Here, $n_\mathbb{a}$ labels an eigenvalue of the number operator $N_\mathbb{a} = \hat{\mathbb{a}}^\dagger \hat{\mathbb{a}}$ and likewise for $n_\mathbb{b}.$ An alternative way to label states is by the total number $N = n_\mathbb{a} + n_\mathbb{b}$ and angular momentum $M = s(n_\mathbb{a} - n_\mathbb{b})$ \footnote{The same symbols $N$ and $M$ have been used to denote the total mode number and angular momentum for different operators. This might lead to some confusion, but we hope that the reader can disentangle easily based on context}:
\begin{equation}
\label{eq: energies in Bogoliubov}
\begin{split}
     \ket{N,M} &= \frac{\left[\hat{\mathbb{a}}^\dagger\right]^{(N+sM)/2}}{\sqrt{[(N+sM)/2]!}}\frac{\left[\hat{\mathbb{b}}^\dagger\right]^{(N-sM)/2}}{\sqrt{[(N-sM)/2]!}} \ket{0,0},    \\
     E_{N,M}(\kappa_z) &= \frac{\hbar\omega_c \sqrt{1-\lambda^2}}{4} \left(N + 1 \right) +  \frac{\hbar^2}{2m} \left(\kappa_z - \beta M\right)^2.
\end{split}
\end{equation}
In either case, the eigenstates are labeled by \textit{three} quantum numbers $(n_\mathbb{a}, n_\mathbb{b}, \kappa_z)$ (or $(N,M,\kappa_z)$) because we are in three dimensions. This is true no matter what form of the regularization is employed because a Bogoliubov transformation must preserve mode number. Our system presents a two-mode bosonic problem both before and after performing the Bogoliubov transformation.

 The phenomenological parameter $\lambda$ is derivable from an isotropic confining potential. By adding a quadratic potential of the following form to the Hamiltonian,
\begin{equation}
    V_0 = \frac{\hbar^2\beta^2q^2}{2m} \left(x^2 + y^2 \right) ,
\end{equation}
we show in Appendix \ref{sec: Special-Function Approach} that the spectrum takes exactly the form of Eq. \eqref{eq: energies in Bogoliubov} with  
\begin{equation}
    \lambda = \sqrt{1-\left(\frac{2q}{\kappa_z}\right)^2},
\end{equation}
where $q^{-1}$ is the confining length scale. As $q \rightarrow 0,$ the confining trap becomes vanishingly weak and the regularization parameter approaches its singular limit of $\lambda = 1.$ In terms of the confining potential, the energy spectrum takes a simple form
\begin{equation}
    E_{N,M}(\kappa_z) = \frac{\hbar^2\beta q}{m} \left(N + 1 \right) +  \frac{\hbar^2}{2m} \left(\kappa_z - \beta M\right)^2,
\end{equation}
which, again, shows that the dispersion in $N$ vanishes in the singular limit as $q \rightarrow0.$ Although not mathematically well-defined, the $q=0$ limit is shown in Fig. \ref{fig: band structure}(b) as dashed lines where all branches of varying $N$ with fixed $M$ collapse into single parabolic bands.

Before the Bogoliubov transformation, Eq. \eqref{eq: kinetic energy in operator form} demonstrates the essential role of the centrifugal potential. In the original basis  the kinetic energy operator contains number-nonconserving terms $\hat a^\dag \hat b^\dag + \hat a \hat b$ that conserve the orbital angular momentum $M$ while coupling different number states $N$ along separate diagonal strings in the Fock space (represented with the bold blue arrows in Fig. \ref{fig: band structure}). Bogoliubov transformations in each string identify the new ground states as superposition states where the total number of quanta $N$ is allowed to fluctuate, allowing radial excursions of the amplitudes  where the energy penalty for promotion to the higher $N$ states is balanced by the energy gained in the centrifugal potential. A representative spectrum of states in the presence of a harmonic confining potential is shown as the bold curves in Fig. \ref{fig: band structure}(b).

\begin{figure*}[ht!]
\centering
\includegraphics[width=\textwidth]{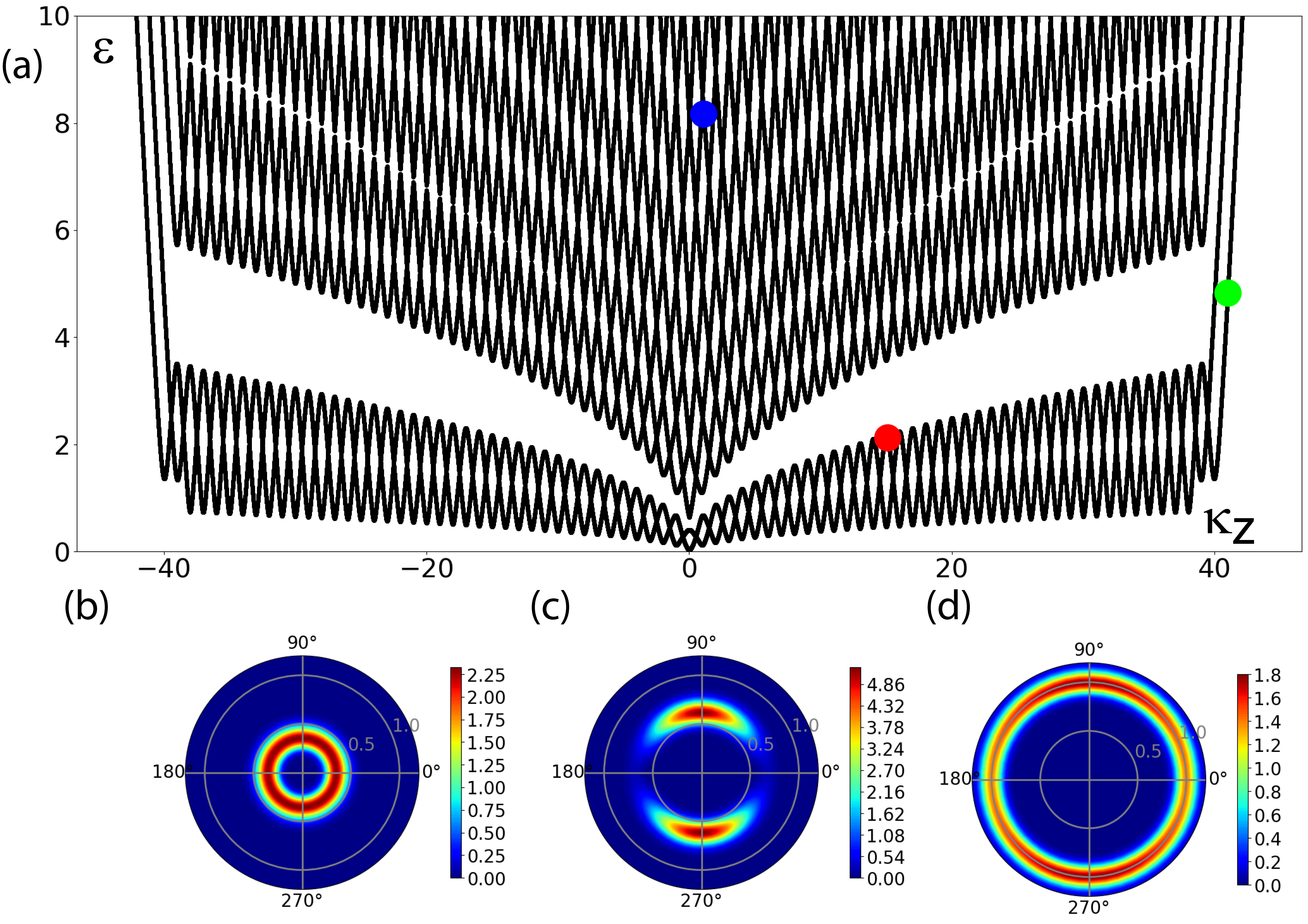}
\caption{(a) Dispersion $\varepsilon(\kappa_z)$ of the Bogoliubov eigenvalues in the presence  a  saddle point potential with discrete $C_2$ symmetry with $\beta = 1$ and $|M| \le 40$ and stabilized by a weak harmonic  potential. In the bulk there is a spectral separation between two branches pinned near the saddle point and a high energy manifold of nearly free weakly scattered states. (b-d) Spatial distributions of the density in Bogoliubov eigenstates at designated places in the spectrum indicated by colored symbols in (a). (b) Mode density in the dispersive bulk spectrum (blue point in (a)). (c) Mode density for a representative solution scattered by the saddle point potential (red point in (a)). (d) Edge state on the outer boundary of the model (green point in (a)).  
\label{fig: band structure 2}}
\end{figure*}

\section{Mixing at a  Saddle Point Potential}
The  representation developed in the previous section provides a useful way to analyze the coherent superpositions of propagating states that occur by scattering these states in a crystal potential with a screw symmetry. Note that in an isotropic medium the level crossings at $\varepsilon \neq 0$ in Fig. \ref{fig: band structure}(b) are protected by angular momentum conservation.  However, a twisted crystal potential lowers this to a discrete rotational symmetry that preserves the screw symmetry. In a typical case the relevant potential has the lateral period of an emergent moir$\acute{\rm e}$ superlattice \cite{andrei_marvels_2021,bistritzer_moire_2011}  but the physical consequences are  captured by an even simpler situation with state mixing occuring at an {\it isolated} saddle point in the potential in real space where the principal axes of the saddle precess with the  pitch of the crystal $\beta$: 
\begin{equation}
V_2  = \frac{k_2}{2} \left( (x-iy)^2 e^{2 i \beta z} + (x+iy)^2 e^{-2i \beta z} \right).
\end{equation}
This potential has twofold rotational symmetry but preserves the continuous screw symmetry. Note that $V_2$ allows mixing of angular momentum branches $M$ and $M \pm 2$ within a common screw sector $\kappa_z$ producing a pattern of avoided crossings.  A useful limiting case is for the extremal states $N=|M|$ where the coherent states have the {\it minimum} allowed radius for each value of the angular momentum $M$ and the saddle potential gaps out the free particle spectrum (Fig. \ref{fig: band structure}(c)) \cite{crepel_microscopic_2020}.  Explicitly, the action of $V_2$ on the number states is
\begin{equation}
    \hat V_2 = - \frac{k_2}{2|\kappa_z \beta|} \, \left( (\hat a^\dag -\hat b)^2 + (\hat a - \hat b^\dag)^2 \right),
\end{equation}
which has the form of a two-mode state-squeezing operator \cite{crepel_geometric_2023}. Because of the quadrupolar form, the scattered states  exchange two units of angular momentum with the twisted lattice  and therefore the angular momentum is conserved mod 2. This isolates two sets of low energy bands in which  the superposed angular momentum states are solely even- or odd- integer valued. The saddle point potential also acts to destabilize the lowest frequency modes in each angular momentum sector (shown as Bogoliubov branches that develop imaginary eigenvalues shown as the red lines in Fig. \ref{fig: band structure}(c) ). This is the signature of unstable trajectories that fall off the saddle point \cite{crepel_geometric_2023}.  This instability could be regulated by introducing quartic terms in the crystal potential which would lead to a positive definite excitation spectrum around a stable state at a shifted equilibrium position.\\

\begin{figure}[ht!]
\centering
\includegraphics[width=85mm]{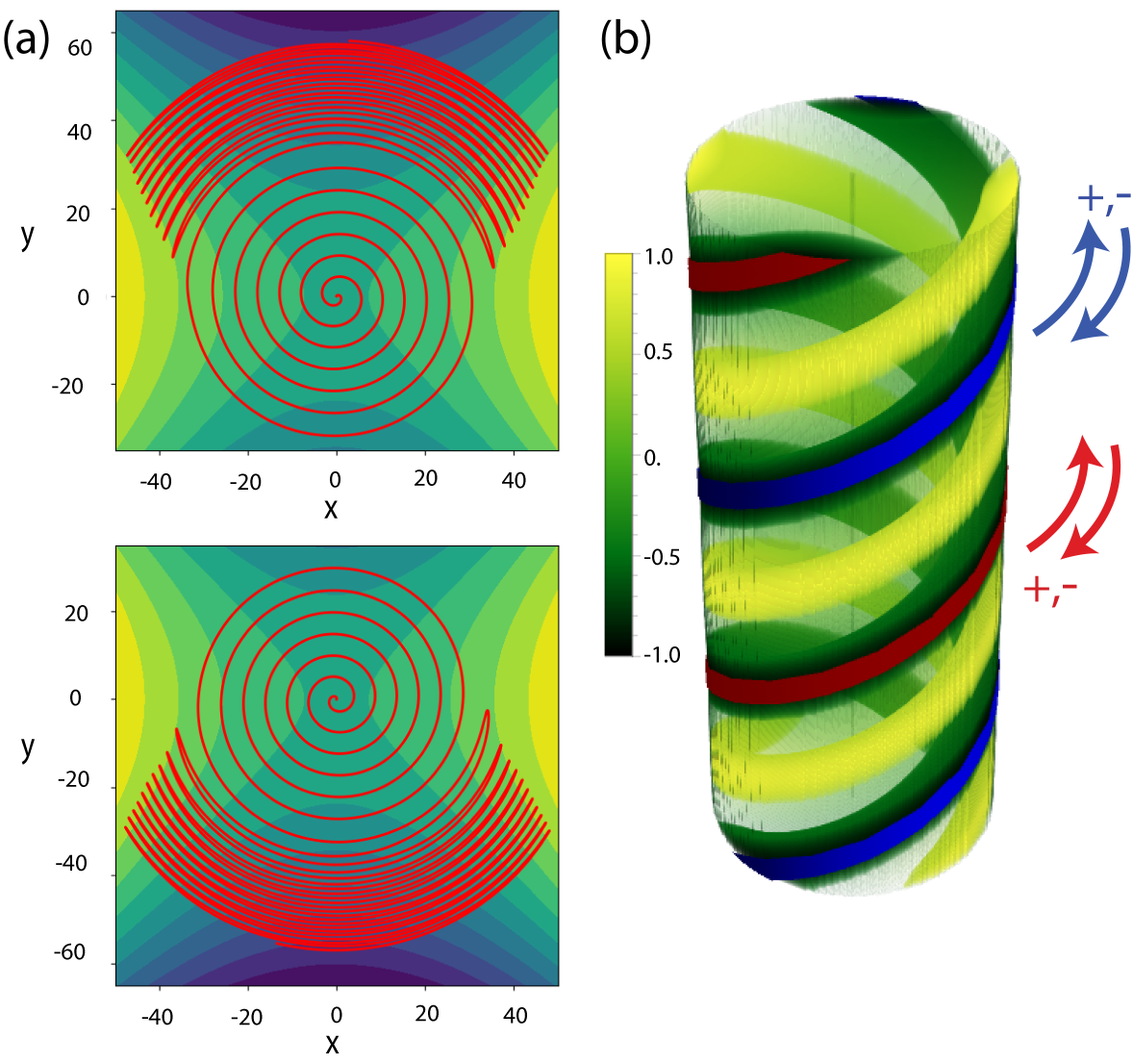}
\caption{Classical solutions for modes in a twisted potential: (a) Classical trajectory of a coherent state released from a saddle point potential in sectors $\pm \kappa_z$ with equipotentials given as a density plot in background. Initially the two orbits circulate in opposite directions before falling off the saddle point.  (b) Density plot: saddle point energy as a function of position in a cylinder. Red and blue lines: surface channels that minimize the  potential energy. In the quantum problem these become helical channels that reside on opposite surfaces in which opposite angular momentum states $(+,-)$ counterpropagate in opposite directions. Each broad band contains two modes on each surface arising from the even- and odd- orbital angular momentum sectors in the bulk.  \label{fig: semiclassical}}
\end{figure}

With open boundary conditions, the radius $R$ of the system sets a bound on the maximum accessible angular momentum states $|M| \le |\kappa_z \beta| \, R^2$. Dispersing branches that saturate this bound occur at the outer boundaries and have no partners at larger radius with which to mix. Consequently these boundary  branches are not gapped out and traverse the bulk gaps as shown in Fig. \ref{fig: band structure}(c) and more clearly in Fig. \ref{fig: band structure 2}(a).  Accounting for the independent even- and odd- integer angular momentum families of low energy modes in the bulk, there are two such helical channels at the edge, each containing counterpropagating modes in which opposite angular momenta propagate ballistically in opposite directions along the screw axis. Combinations of these edge modes from the even- and odd- angular momentum families are ``barber pole" channels forming pairs of winding trajectories around the boundary.  Backscattering in the surface channels is strongly suppressed because of the large mismatch of their screw eigenvalues $\pm \kappa_z$  (though in principle it would be symmetry allowed for sufficiently large momentum scattering), analogous to the symmetry protection of surface modes on topological crystalline insulators \cite{fu_topological_2011, kruthoff_topological_2017}.

\section{Semiclassical Model}

Some features of these quantum states can  be understood from classical considerations and are expected to occur in any twisted medium.  To simulate the semiclassical dynamics, we numerically integrate the following equations of motion:
\begin{equation}
\begin{split}
    m \ddot{\mathbf{r}}(t) &= -\nabla_\mathbf{r} V_2(\mathbf{r}) + q \dot{\mathbf{r}} \times \mathbf{B}, \\
    V_2(\mathbf{r}) &= k_2( x^2 - y^2 ), \\
    \mathbf{r}(t=0) & = \mathbf{0}, \quad \dot{\mathbf{r}}(t = 0) = \mathbf{v}_0 \neq \mathbf{0}.
\end{split}
\end{equation} In Fig. \ref{fig: semiclassical} (a) we display the solution of the classical equation of motion giving the time evolution of the mean position of a coherent state released from the saddle point with a small initial velocity. We observe trajectories in the form of  circulating orbits whose direction is determined by the sign of $\kappa_z \beta$ and which grow in size drifting off the saddle point potential before ultimately repeatedly reflecting at turning points on the sidewalls of the saddle potential.

The structure of the edge states also follows from a classical model. If we null the kinetic energy, the state of minimum potential energy occurs along the unstable directions of the saddle at the sample boundary (Fig. \ref{fig: semiclassical}(b)). On a cylinder these extrema of the saddle potential  form a pair of classical winding barber pole channels on opposite surfaces. The even and odd superpositions of these surface modes  correspond to the pairs of edge states found in the even- and odd- angular momenta surface channels in the quantum theory.

Removing the $N=|M|$ restriction allows the radii of the coherent states to adjust on the saddle potential. Including these processes we again find that the two lowest branches of the coupled spectrum separate from  a manifold of higher energy states and generate a pair of helical edge states at the outer surface that traverse this gap (see Appendix \ref{sec: Boundary modes of Hofstadter networks}).  For the general case with $N \neq |M|$ the spectrum contains  these fundamental modes along with a pattern of overtones which are quantized excitations of the orbitals from their optimized shapes in the confining potential. These  overtones form narrow bands in the bulk each of which evolve into one-sided  branches that are distributed beneath the surface forming a ``soft edge" at reduced radii $R$.  The lowest energy such branches are the fundamental edge states pinned to the exterior boundary described above in the ``stiff-orbit" model shown in Fig. \ref{fig: band structure 2}.

\section{Hofstadter Network}

\begin{figure*}[ht!]
\centering
\includegraphics[width=170mm]{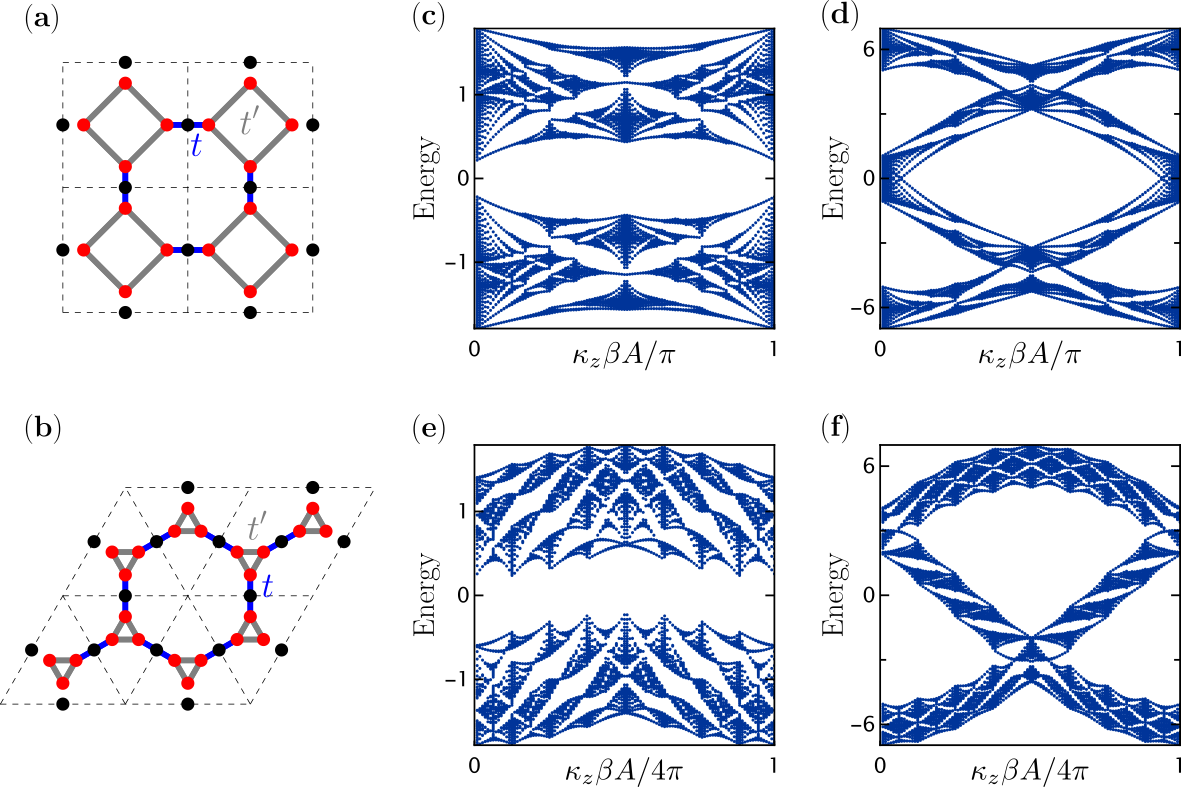}
\caption{Hofstadter networks for dispersive low energy bands in a three-dimensional twisted crystal. (a,b) Networks link stable equilibrium points (red) near the saddle points (black) in the superlattice potential. (c,d) Hofstadter spectra for network models on a square Bravais lattice with effective flux $2 \hbar\kappa_z \beta A/e$ per unit cell with area $A$.  (e,f) Hofstadter spectra for network models on a triangular lattice with  effective flux  $2 \hbar \kappa_z \beta A/e$ per unit cell with area $A$. In (c) and (e) the intra-saddle point tunneling amplitude  (blue links) $|t|=1$ and the inter saddle point amplitude (grey links) $ t' = - 0.4$. In (d) and (f) $|t| =1, \, t' = -3$. \label{fig: Hofstadter}}
\end{figure*}

In a three-dimensional twisted crystal each layer sees a potential that is periodic on an emergent moir$\acute{\rm e}$ superlattice scale \cite{andrei_marvels_2021}.  In this situation the single saddle point (SP) is replaced by periodic energy landscape containing a translationally ordered array of compensating SP's located at sublattice positions  on a Bravais superlattice (Fig. \ref{fig: Hofstadter}(a,b)).  We study this in a model where an intralayer scalar potential on the moir$\acute{\rm e}$ length scale is induced by the mismatch of translation vectors on the nearest layer neighbor layers.  

In the strong field limit, the Coriolis deflection in such a scalar potential confines states in  closed orbits near the potential extrema.  Unstable  trajectories that propagate along paths that connect SP's can escape by propagating along  the links of two-dimensional periodic networks \cite{fertig_transmission_1987,chalker_percolation_1988,pal_emergent_2019,de_beule_network_2021}.  Importantly, although the crystal potential is invariant under the network lattice translations $\{ {\bm R}\}$, the coherent states accumulate path-dependent phases  when displaced by $ {\bm R}$'s  making them noncommutative \cite{zak_magnetic_1964,haldane_origin_2018}. Thus the superlattice displacements are generally {\it not}  identified with the set of magnetic translations $\{ {\bm T} \}$ \cite{zak_magnetic_1964} that define an effective magnetic unit cell with area $\propto 1/2 |\kappa_z \beta|$ which is continuously tunable by the twist eigenvalue. This noncommutative structure of the elementary translations provides a {\it local} measure for topological dynamics within in a projected Hilbert space even in the absence of long range crystalline order \cite{bianco_mapping_2011,dereli_bloch_2021}.  Note that here the effective magnetic flux that links a single unit cell of the moir$\acute{\rm e}$ superlattice  changes continuously as a function of the conserved twist momentum $\kappa_z$ and this produces quantum interference in a Hofstadter spectrum whose gap structure depends on the connectivity of the network. Two examples are displayed in Fig. \ref{fig: Hofstadter} for models constructed on square and triangular Bravais lattices. In either case the structure of the Hofstadter spectra depend on whether the intra-saddle point amplitudes (double well tunneling between a pair of equilibrium states bound near a single SP \cite{novaes_generalized_2003}) or inter-saddle point amplitude \cite{fertig_transmission_1987,chalker_percolation_1988} (paths connecting neighboring SP's) are dominant. In the latter case the spectrum shows the signature of well developed resonances from weakly coupled intra-cell orbital loops that are linked by the effective flux. 

Either limit in Fig. \ref{fig: Hofstadter} should be physically realizable, being controlled by the relative strengths of the  moir$\acute{\rm e}$ crystal field, the kinetic energy  penalty for confining states within a  moir$\acute{\rm e}$ cell and the strength of the regularizing (pinning) potential.  In a typical case, for sufficiently small rotation angle and weak pinning potential, the resonant patterns in Fig. \ref{fig: Hofstadter}(d) and \ref{fig: Hofstadter}(f) should be generic.  They describe a reconstruction of the low energy degrees of freedom for a nearly free particle in a twisted crystal potential that produces orbital degeneracies on the sublattice nodes of an emergent  moir$\acute{\rm e}$ superlattice. Note that the edge state structure for the single SP problem (Fig. \ref{fig: band structure 2}) reconstructs in the super moir$\acute{\rm e}$ problem so that boundary modes appear within the bulk Hofstadter gaps (Fig. \ref{fig: Hofstadter})  (see Appendix \ref{sec: Boundary modes of Hofstadter networks}). 


\section{Discussion}

The twist-derived gauge coupling in this model is time-reversal symmetric and by itself does not produce a Hall response: opposing contributions to response functions from twist sectors $\pm \kappa_z$ cancel. This can be avoided by biasing the populations of the orbital states with an applied magnetic field, by exciting with circularly polarized light \cite{ji_opto-twistronic_2023} or by spontaneous symmetry breaking from the projected interactions between electrons residing in these narrow bands. 

Many of the features found here in three dimensions have striking counterparts  in twisted {\it two-dimensional} few-layer van der Waals materials \cite{andrei_marvels_2021}. Theory suggests and experiments observe bands of low energy degrees of freedom, possibly with topological character, spectrally isolated from remote manifolds of higher energy dispersive states. The conditions under which these low energy bands can develop coherence with the analytic structure of lowest Landau level states has been addressed  in  recent studies \cite{tarnopolsky_origin_2019,ledwith_vortexability_2023}.  A promising approach is to now relate the ideas developed here in three dimensions to the phenomenology of quasi two-dimensional ultrathin materials.

\section*{Acknowledgements}
We thank R. Kamien and C. Kane for helpful discussions and V. Cr$\acute{\rm e}$pel and R. Fletcher for useful correspondence. This work was supported by the Department of Energy under grant DE-FG02-84ER45118, Royal Society exchange grant IES/R1/221060 and Singapore Ministry of Education AcRF Tier 2 Grant No. MOE-T2EP50220-0016. V.T.P. is supported by C. Lewandowski's start-up funds from Florida State University and the National High Magnetic Field Laboratory. The National High Magnetic Field Laboratory is supported by the National Science Foundation through NSF/DMR-2128556 and the State of Florida.

\appendix

\onecolumngrid

\section{A Tale of Two Reference Frames}

\subsection{Coordinate Systems}

\begin{figure}[h!]
    \centering
    \includegraphics[width=0.4\linewidth]{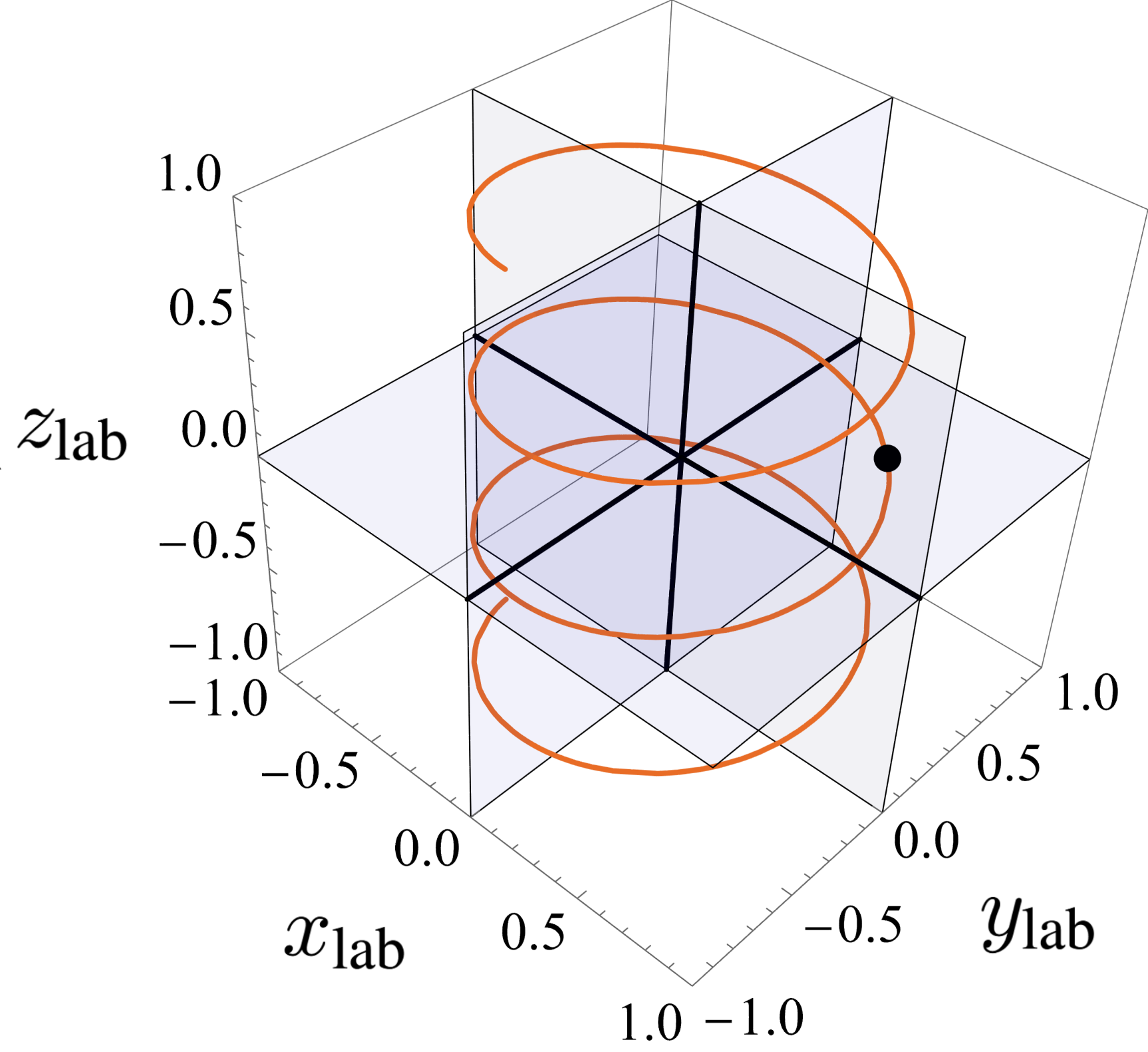}
    \caption{\textbf{Helical coordinate system.} Points of fixed $(x_\text{twist}, y_\text{twist}) = (1/2,1/2)$ are plotted in orange. }
    \label{fig:coordinate}
\end{figure}

In a three-dimensional twisted crystal, there are two equivalent frames of reference using which to describe the dynamics. The first is the usual laboratory frame. This is the frame adopted in the main text because it is familiar and intuitive. It takes the perspective of an external observer. In this frame, we can either affix a rectangular coordinate system $(x_\text{lab}, y_\text{lab}, z_\text{lab})$ or a cylindrical coordinate system $(r_\text{lab}, \phi_\text{lab}, z_\text{lab}).$ It seems only sensible to place the origin at the twist axis.  The second is the crystal frame, also referred to as the twisted frame, in which the coordinates co-rotate with the crystal. This frame adopts the point of view of the crystal instead of that of the external observer; therefore, it   naturally encodes the helical symmetry of the crystal, and consequently, allows certain crystal potentials to be written in a simplified manner. In this frame, we can also affix a helical coordinate system which is defined relative to the laboratory coordinates as follows \cite{vachagina2014use}:
\begin{equation}
    \begin{pmatrix}
        x_\text{twist} \\
        y_\text{twist} \\
        z_\text{twist}
    \end{pmatrix} =   \begin{pmatrix}
        \cos \left( \beta z_\text{lab} \right) & \sin \left( \beta z_\text{lab} \right) & 0 \\
        -\sin \left( \beta z_\text{lab} \right) & \cos \left( \beta z_\text{lab} \right) & 0 \\
        0 & 0 & 1
    \end{pmatrix}  \begin{pmatrix}
        x_\text{lab} \\
        y_\text{lab} \\
        z_\text{lab}
    \end{pmatrix},
\end{equation}
where $\beta$ is the pitch of the twist (and has dimension of inverse length). This frame is chosen such that a fixed $(x_\text{twist}, y_\text{twist})$ point winds in counterclockwise fashion as $z_\text{twist} = z_\text{lab}$ increases, as shown in Fig. \ref{fig:coordinate}.  In the laboratory frame, the Laplacian operator is simply
\begin{equation}
    \Delta = \frac{\partial^2}{\partial x_\text{lab}^2} + \frac{\partial^2}{\partial y_\text{lab}^2} + \frac{\partial^2}{\partial z_\text{lab}^2}.
\end{equation}
In the twisted frame, the Laplacian operator is more complicated
\begin{equation}
    \Delta = \frac{\partial^2}{\partial x_\text{twist}^2} + \frac{\partial^2}{\partial y_\text{twist}^2} + \left( \frac{\partial}{\partial z_\text{twist}} - \beta \left[ x_\text{twist} \frac{\partial}{\partial y_\text{twist}}  - y_\text{twist} \frac{\partial}{\partial x_\text{twist}} \right] \right)^2.
\end{equation}
This can be derived straightforwardly using repeated applications of the chain rule: we observe that the in-plane component of the Laplacian has the same form in both frames, $\partial^2_{ x_\text{lab}} + \partial^2_{ y_\text{lab}} = \partial^2_{ x_\text{twist}} + \partial^2_{ y_\text{twist}} ,$ while the out-of-plane component of the Laplacian is more complicated in the twisted frame, $\partial^2_{z_\text{lab}} = \left( \partial_{z_\text{twist}} - \beta \left[ x_\text{twist} \partial_{y_\text{twist}}  - y_\text{twist} \partial_{x_\text{twist}} \right] \right)^2.$ Here, a term related to the angular momentum operator, $x_\text{twist} \partial_{ y_\text{twist}}  - y_\text{twist} \partial_{x_\text{twist}},$ naturally makes its appearance. The integration measure in both frames is of the same form: $dx_\text{twist}dy_\text{twist}dz_\text{twist}=dx_\text{lab}dy_\text{lab}dz_\text{lab}.$

\subsection{Dual Representations of the Free-Particle Hamiltonian}
\label{sec: Dual Representations of the Free-Particle Hamiltonian}

In this work, we consider only Schrodinger particles with quadratic dispersion. In any frame, the free-particle Hamiltonian is just $\hat{\mathcal{K}} = \hat{\mathbf{p}} \cdot \hat{\mathbf{p}}/2m,$ where $m$ is the mass of the particle. In the laboratory frame, the momentum operator is 
\begin{equation}
    \left( \hat{p}_{x_\text{lab}} , \hat{p}_{y_\text{lab}} , \hat{p}_{z_\text{lab}} \right) = \frac{\hbar}{i} \left( \frac{\partial}{\partial x_\text{lab}} , \frac{\partial}{\partial y_\text{lab}} , \frac{\partial}{\partial z_\text{lab}}   \right).
\end{equation}
Thus, the Hamiltonian in the laboratory frame has the representation
\begin{equation}
\label{eq: Hamiltoniani in laboratory frame}
    \hat{\mathcal{K}} = - \frac{\hbar^2}{2m} \left[ \frac{\partial^2}{\partial x_\text{lab}^2} + \frac{\partial^2}{\partial y_\text{lab}^2} + \frac{\partial^2}{\partial z_\text{lab}^2} \right].
\end{equation}
Similarly, the Hamiltonian in the twisted frame has the representation
\begin{equation}
\begin{split}
\label{eq: Hamiltoniani in twisted frame}
    \hat{\mathcal{K}} &= - \frac{\hbar^2}{2m} \left[ \frac{\partial^2}{\partial x_\text{twist}^2} + \frac{\partial^2}{\partial y_\text{twist}^2} + \left( \frac{\partial}{\partial z_\text{twist}} - \beta \left[ x_\text{twist} \frac{\partial}{\partial y_\text{twist}}  - y_\text{twist} \frac{\partial}{\partial x_\text{twist}} \right] \right)^2 \right] \\
    &= \frac{1}{2m} \left[ \hat{p}^2_{x_\text{twist}} + \hat{p}^2_{y_\text{twist}} + \left[\hat{p}_{z_\text{twist}} - \beta \left( \hat{x}_\text{twist} \hat{p}_{y_\text{twist}} - \hat{y}_\text{twist} \hat{p}_{x_\text{twist}} \right) \right]^2  \right],
\end{split}
\end{equation}
where we have defined new momentum operators in the twisted frame analogously to those defined in the laboratory frame. It is apparent that $\hat{p}_{z_\text{twist}}$ commutes with the Hamiltonian, so we can project into a sector of definite eigenvalue $\hbar k_z$ of $\hat{p}_{z_\text{twist}}.$ This allows us to replace $\hat{p}_{z_\text{twist}}$ with a $c$-number: $2m \mathcal{K} =  \hat{p}^2_{x_\text{twist}} + \hat{p}^2_{y_\text{twist}} + \left[\hbar k_z - \beta \left( \hat{x}_\text{twist} \hat{p}_{y_\text{twist}} - \hat{y}_\text{twist} \hat{p}_{x_\text{twist}} \right) \right]^2  .$ To bring Eq. \eqref{eq: Hamiltoniani in laboratory frame} into a form similar to Eq. \eqref{eq: Hamiltoniani in twisted frame}, we define a screw operator in the laboratory frame:
\begin{equation}
    \hbar\hat{\kappa}_z = \hat{p}_{z_\text{lab}} + \beta \left( \hat{x}_\text{lab} \hat{p}_{y_\text{lab}} - \hat{y}_\text{lab} \hat{p}_{x_\text{lab}} \right).
\end{equation}
In terms of the screw operator, we have $2m\hat{\mathcal{K}} = \hat{p}^2_{x_\text{lab}} + \hat{p}^2_{y_\text{lab}} + \left[\hbar\hat{\kappa}_z - \beta \left( \hat{x}_\text{lab} \hat{p}_{y_\text{lab}} - \hat{y}_\text{lab} \hat{p}_{x_\text{lab}} \right) \right]^2.$ Using the canonical commutation relations $\left[ \hat{x}_{i,\text{lab}} , \hat{p}_{x_{j,\text{lab}}} \right] = i \hbar \delta_{ij},$ we find that 
\begin{equation}
    \begin{split}
        \left[ \hat{p}_{x_\text{lab}}, \hbar \hat{\kappa}_z \right] &= \left[ \hat{p}_{x_\text{lab}}, \beta  \hat{x}_\text{lab} \hat{p}_{y_\text{lab}}\right] =  -i \beta \hbar \hat{p}_{y_\text{lab}} \rightarrow \left[ \hat{p}^2_{x_\text{lab}}, \hbar \hat{\kappa}_z \right] =  -2i \beta \hbar \hat{p}_{y_\text{lab}} \hat{p}_{x_\text{lab}}\\
        \left[ \hat{p}_{y_\text{lab}}, \hbar \hat{\kappa}_z \right] & = -\left[ \hat{p}_{y_\text{lab}}, \beta  \hat{y}_\text{lab} \hat{p}_{x_\text{lab}} \right]  =  i \beta \hbar \hat{p}_{x_\text{lab}} \rightarrow \left[ \hat{p}^2_{y_\text{lab}}, \hbar \hat{\kappa}_z \right] =  +2i \beta \hbar \hat{p}_{y_\text{lab}} \hat{p}_{x_\text{lab}}, \\ 
        \left[ \hat{p}_{z_\text{lab}}, \hbar \hat{\kappa}_z \right] & = 0 \rightarrow \left[ \hat{p}^2_{z_\text{lab}}, \hbar \hat{\kappa}_z \right] = 0.
    \end{split}
\end{equation}
Therefore, $\left[ \hat{\mathcal{K}}, \hbar \hat{\kappa}_z \right] = \left[(\hat{p}^2_{x_\text{lab}} + \hat{p}^2_{y_\text{lab}} + \hat{p}^2_{z_\text{lab}})/2m, \hbar \hat{\kappa}_z \right] = 0. $ Thus, we can also project into subspaces of definite screw eigenvalues $\kappa_z$ of $\hat{\kappa}_z.$ In the end, we have two different representations of the same Hamiltonian with the exact same form:
\begin{equation}
    \begin{split}
        \hat{\mathcal{K}}(k_z) &= \frac{1}{2m} \left[ \hat{p}^2_{x_\text{twist}} + \hat{p}^2_{y_\text{twist}} + \left[\hbar k_z - \beta \left( \hat{x}_\text{twist} \hat{p}_{y_\text{twist}} - \hat{y}_\text{twist} \hat{p}_{x_\text{twist}} \right) \right]^2  \right], \\
        \hat{\mathcal{K}}(\kappa_z) &= \frac{1}{2m} \left[ \hat{p}^2_{x_\text{lab}} + \hat{p}^2_{y_\text{lab}} + \left[\hbar \kappa_z - \beta \left( \hat{x}_\text{lab} \hat{p}_{y_\text{lab}} - \hat{y}_\text{lab} \hat{p}_{x_\text{lab}} \right) \right]^2  \right]. \\
    \end{split}
\end{equation}    
Despite apparent formal equivalence, the reader is reminded that these two representations are in two completely different frames, and consequently are projected into different physical sectors: in the twisted frame, the Hamiltonian is projected into sectors with plane wavevectors $k_z$, while in the laboratory frame, the Hamiltonian is projected into sectors with definite screw eigenvalues $\kappa_z$. Nonetheless, the formal equivalence makes it explicit that both frames are equally useful as long as appropriate care is dedicated when physical interpretation is required. For formal manipulations, only commutation relations matter, and so, both representations give identical results. Consequently, we shall drop the subscripts indicating frames henceforth for brevity whenever there is no risk for confusion (restoring them only when necessary for emphasis), and the reader is free to choose either $k_z$ or $\kappa_z$ to do the projection. In keeping with the main text, we shall choose $\kappa_z$ going forward.

\subsection{Plane Waves versus Screw States}

In the twisted frame, we use $k_z$ to label eigenvalues of the operator $\hat{p}_{z_\text{twist}} = -i\hbar\partial_{z_\text{twist}}.$ The associated eigenfunctions are simply plane waves $e^{ik_z z_\text{twist}} .$  In the laboratory frame, we use $\kappa_z$ to label eigenvalues of the operator $\hat{\kappa}_z = -i\partial_{ z_\text{lab}} -i \beta \left( x_\text{lab} \partial_{ y_\text{lab}}  - y_\text{lab} \partial_{ x_\text{lab}} \right).$ Actually, $\hbar\hat{\kappa}_z$ and $\hat{p}_{z_\text{twist}}$ are the same operator just expressed in two different frames; the use of $k_z$ and $\kappa_z$ here is to emphasize which frame we are using. In polar coordinates, $x_\text{lab} = r_\text{lab} \cos \phi_\text{lab}$ and $y_\text{lab} = r_\text{lab} \sin \phi_\text{lab},$ this operator is just
\begin{equation}
    \hat{\kappa}_z = -i\partial_{ z_\text{lab}}  -i \beta \partial_{ \phi_\text{lab}} .
\end{equation}
It is then straightforward to observe that the eigenfunctions of $\hat{\kappa}_z$ are of the form
\begin{equation}
    \exp \left[ i \left( \kappa_z - \beta M \right) z_\text{lab} + i M \phi_\text{lab}  \right],
\end{equation}
where $M$ is an integer labeling the angular momentum. Rewritten in the twisted frame where $\phi_\text{twist} = \phi_\text{lab} - \beta z_\text{lab},$ these eigenvalues have the nice form $\exp \left[ i \kappa_z z_\text{twist} + i M \phi_\text{twist}  \right]$ (notice that this is also an eigenfunction of $\hat{p}_{z_\text{twist}}$  with eigenvalue $\kappa_z$). We notice that the screw states differ from the plane waves only in that they encode a helical twist explicitly. Because of that, in the laboratory frame, these states depend on both $z$ and $\phi$, and consequently, they depend also on angular momentum. In other words, eigenstates of $\hat{\kappa}_z$ couple the in-plane and out-of-plane degrees of freedom (unless the angular momentum is zero). Said more simply, in the twisted frame, the eigenvalue of $\hat{p}_{z_\mathrm{twist}}$ is just the coefficient in front of $z_\mathrm{twist}$ in the exponential, i.e. the wavenumber; in the laboratory frame,  the eigenvalue of $\hat{\kappa}_z$ is \textit{not} just the coefficient in front of $z_\mathrm{lab}$ in the exponential but the eigenvalue also depends on the angular momentum $M$.

\section{Landau-Level Representation of the Free-Particle Hamiltonian}
\label{sec: kappaz = 0}

In Sec. \ref{sec: Landau-Level Representation of the Free-Particle Hamiltonian} of the  main text, we have presented the Landau-level representation of the Hamiltonian when $\beta \kappa_z \neq 0$. For completeness, we now handle the $\beta \kappa_z = 0$ case. We always assume $\beta \neq 0$ for otherwise there is no twist. The Hamiltonian now is just $\hat{\mathcal{K}} = \frac{1}{2m} \left[ \hat{p}^2_{x} + \hat{p}^2_{y} + \beta^2 \left( \hat{x} \hat{p}_{y} - \hat{y} \hat{p}_{x} \right)^2 \right].$ As mentioned previously, the operators defined before degenerate $\hat{\pi}_i = \hat{\kappa}_i = \hat{p}_i,$ and the ladder operators $\hat{a}$ and $\hat{b}$ no longer couple to the coordinates $\hat{x}$ and $\hat{y}.$ So this construction does not work for the $\kappa_z = 0$ case.  However, a very similar construction does work. Essentially, we define a new operator $\hat{a}_0$ analogous to $\hat{a}$ with $\beta \kappa_z$ replaced by just $\beta^2$ and a new operator $\hat{b}_0$ analogous to $\hat{b}$ also with $\beta \kappa_z$ replaced by just $\beta^2$ (this replacement is permissible since $\beta$ has units of $\kappa_z$)
\begin{equation}
\label{eq: equivalence between afrak bfrak and x y}
    \begin{pmatrix}
        \hat{a}_0 \\
        \hat{a}_0^\dagger \\
        \hat{b}_0 \\
        \hat{b}_0^\dagger
    \end{pmatrix} = \frac{\ell_0}{\sqrt{2}\hbar}\begin{pmatrix}
        1 & -i &-i & -1 \\
         1 & i &i & -1 \\
         1 & -i & i &  1 \\
         1 & i & -i &  1
    \end{pmatrix} \begin{pmatrix}
        \hat{p}_x \\
         \hbar\beta^2\hat{x}\\
        \hat{p}_y \\
         \hbar\beta^2\hat{y}
    \end{pmatrix} \rightarrow  \begin{pmatrix}
        \hat{p}_x \\
        \hbar\beta^2\hat{x}\\
        \hat{p}_y \\
        \hbar\beta^2\hat{y}
    \end{pmatrix} = \frac{\hbar}{2\sqrt{2}\ell_0}\left(
\begin{array}{cccc}
 1 & 1 & 1 & 1 \\
 i  & -i  & i & -i \\
 i & -i  & -i & i\\
 -1 & -1 & 1 & 1 \\
\end{array}
\right)\begin{pmatrix}
        \hat{a}_0 \\
        \hat{a}_0^\dagger \\
        \hat{b}_0 \\
        \hat{b}_0^\dagger
    \end{pmatrix},
\end{equation}
where $\ell_0^2 = (2\beta^2)^{-1}.$ We inspect the commutation relations: $\left[\hat{a}_0, \hat{a}_0^\dagger\right] =  \left[\hat{b}_0, \hat{b}_0^\dagger\right] = 1$ and $\left[\hat{a}_0, \hat{b}_0^\dagger\right] = \left[\hat{a}_0, \hat{b}_0\right] =0.$ Using these relations, we find
\begin{equation}
\begin{split}
    \hat{p}_x^2 + \hat{p}_y^2 &= \frac{\hbar^2}{2\ell_0^2} \left( \hat{a}_0^\dagger + \hat{b}_0 \right)\left( \hat{a}_0 + \hat{b}_0^\dagger \right),\\
    \hat{x} \hat{p}_y - \hat{y} \hat{p}_x &= \hbar \left( \hat{a}_0^\dagger \hat{a}_0 - \hat{b}_0^\dagger\hat{b}_0\right).
\end{split}
\end{equation}
Now defining $ \omega_0 = \hbar/m \ell_0^2,$ we can write the Hamiltonian as 
\begin{equation}
\label{eq: nice form of equation 2}
    \hat{\mathcal{K}} = \frac{\hbar \omega_0}{4} \left( \hat{a}_0^\dagger\hat{a}_0 + \hat{b}_0^\dagger \hat{b}_0 +\hat{a}_0^\dagger \hat{b}_0^\dagger + \hat{b}_0 \hat{a}_0 + 1  \right) + \frac{\hbar^2\beta^2}{2m}\left( \hat{a}_0^\dagger \hat{a}_0 - \hat{b}_0^\dagger\hat{b}_0\right)^2.
\end{equation}
We note that Eq. \eqref{eq: nice form of equation 2} is  identical to Eq. \eqref{eq: nice form of equation} with $\kappa_z=0$ and the appropriate replacements of operators and constants. This is a nice consistency check. However, it is worth pointing out that while Eq. \eqref{eq: nice form of equation} depends on the sign of $\beta \kappa_z,$ Eq. \eqref{eq: nice form of equation 2} does not depend on the sign of $\beta$ at all since all the $\beta$ factors appear as $\beta^2.$ This is because when $\kappa_z = 0,$ there is no propagation along the vertical direction, so the corresponding eigenstates cannot distinguish between a left-handed twist and a right-handed twist. With that important distinction made, one can now solve Eq. \eqref{eq: nice form of equation 2} using the same techniques used for Eq. \eqref{eq: nice form of equation}. This completes the construction of the Landau-level representation of the free-particle Hamiltonian in all relevant scenarios.

\section{Diagonalization of the (Almost) Free-Particle Hamiltonian}

\subsection{Special-Function Approach}
\label{sec: Special-Function Approach}

There are different ways to diagonalize Eq. \eqref{eq: kinetic energy}. Here, we take the most elementary approach using special functions. It is simplest to work in polar coordinates where the Hamiltonian takes the form
\begin{equation}
    \hat{\mathcal{K}} = -\frac{\hbar^2}{2m} \left[ \frac{\partial^2}{\partial r^2}  + \frac{1}{r} \frac{\partial}{\partial r} + \frac{1}{r^2} \frac{\partial^2}{\partial \phi^2} + \left(i\kappa_z - \beta \frac{\partial}{\partial \phi} \right)^2\right],
\end{equation}
where $x = r \cos \phi$ and $y = r \sin \phi.$ One can choose to be in either the twisted frame or laboratory frame according to the discussion in Sec. \ref{sec: Dual Representations of the Free-Particle Hamiltonian}. The permissible eigenfunctions and energies are
\begin{equation}
\begin{split}
    \psi_{\kappa_z \in \mathbb{R}, M \in \mathbb{Z}, \kappa_r \in \mathbb{R}_{>0}}(r,\phi, z) &=  \left[\frac{\pi \kappa_r^2}{4L^2 A}\right]^{\frac{1}{4}}e^{i\left(\kappa_z-\beta M \right) z + i M \phi} J_{M}(\kappa_r r), \\
    \psi_{\kappa_z  \in \mathbb{R} , M=0, \kappa_r=0}(r,\phi, z) &=  \frac{1}{\sqrt{LA}} e^{i \kappa_z z }, \\
    E(\kappa_z, M, \kappa_r) &= \frac{\hbar^2}{2m} \left[ \kappa_r^2 + \left( \kappa_z -\beta M \right)^2 \right],
\end{split}
\end{equation}
where $L$ and $A$ are the vertical and lateral sizes of the box regularization employed to normalize the eigenstates, and $J_M$ is a Bessel function of the first kind. It is important to note that these functions are only \textit{marginally normalizable} (there are other solutions but those are not even marginally normalizable). This makes sense since we are dealing with a free particle without any confining potential. These solutions are known as Bessel beams in the context of electron vortex states \cite{schattschneider2011theory}.

The Bessel-function basis is qualitatively very different from the basis states of the operators defined in Sec. \ref{sec: Landau-Level Representation of the Free-Particle Hamiltonian}. While the former is extended, the latter is exponentially well-localized. To see this, we observe that basis states in Sec. \ref{sec: Landau-Level Representation of the Free-Particle Hamiltonian} are:
\begin{equation}
    \ket{n_a,n_b} = \frac{\left[\hat{a}^\dagger\right]^{n_a}\left[\hat{b}^\dagger\right]^{n_b}}{\sqrt{n_a!n_b!}}\ket{0,0},
\end{equation}
where $\ket{0,0}$ is the reference state annihilated by both $\hat{a}$ and $\hat{b}.$ In real space, this reference state is just a Gaussian, and it has zero angular momentum: 
\begin{equation}
    \psi_{0,0}(\mathbf{r}) = \frac{1}{\sqrt{2\pi \ell_{\kappa_z}^2L}} \exp \left(- \frac{r^2}{4 \ell_{\kappa_z}^2} + i \kappa_z z \right).
\end{equation}
To find a general state, let us rewrite the operators in modified complex coordinates
\begin{equation}
    \begin{split}
        \hat{a}^\dagger &= \frac{ \ell_{\kappa_z}}{\sqrt{2}} \left(-i  \frac{\partial}{\partial x} + i s \beta \kappa_z x + s  \frac{\partial}{\partial y} -  \beta \kappa_z y\right) = -i\sqrt{2}\ell_{\kappa_z} \left(  \frac{\partial}{\partial \bar{\mathfrak{z}}_s} - \frac{\mathfrak{z}_s}{4\ell_{\kappa_z}^2} \right) = -i \sqrt{2} \ell_{\kappa_z} e^{+ \bar{\mathfrak{z}}_s \mathfrak{z}_s/4\ell_{\kappa_z}^2} \partial_{\bar{\mathfrak{z}}_s}e^{- \bar{\mathfrak{z}}_s \mathfrak{z}_s/4\ell_{\kappa_z}^2}, \\
        \hat{b}^\dagger &= \frac{ \ell_{\kappa_z}}{\sqrt{2}} \left(-i  \frac{\partial}{\partial x} + i s \beta \kappa_z x - s  \frac{\partial}{\partial y} +  \beta \kappa_z y\right) = -i\sqrt{2}\ell_{\kappa_z} \left(  \frac{\partial}{\partial \mathfrak{z}_s} - \frac{\bar{\mathfrak{z}}_s}{4\ell_{\kappa_z}^2} \right)= -i \sqrt{2} \ell_{\kappa_z} e^{+ \bar{\mathfrak{z}}_s \mathfrak{z}_s/4\ell_{\kappa_z}^2} \partial_{\mathfrak{z}_s}e^{- \bar{\mathfrak{z}}_s \mathfrak{z}_s/4\ell_{\kappa_z}^2},
    \end{split}
\end{equation}
where we have adopted a modified convention for the complex coordinates and Wirtinger derivatives:
\begin{equation}
    \begin{split}
        \mathfrak{z}_s &= x + isy, \quad \frac{\partial}{\partial \mathfrak{z}_s} = \frac{1}{2} \left( \frac{\partial}{\partial x} - is \frac{\partial}{\partial y}  \right),  \\
        \bar{\mathfrak{z}}_s &= x - isy, \quad         \frac{\partial}{\partial \bar{\mathfrak{z}}_s} = \frac{1}{2} \left( \frac{\partial}{\partial x} + is \frac{\partial}{\partial y}  \right).
    \end{split}
\end{equation}
We observe the action of these operators on the $\psi_{0,0}(\mathbf{r})$ state:
\begin{equation}
\label{eq: LG wf 1}
    \psi_{n_a,n_b}(\mathbf{r}) = \frac{\left[\hat{a}^\dagger\right]^{n_a}\left[\hat{b}^\dagger\right]^{n_b}}{\sqrt{n_a!n_b!}} \psi_{0,0}(\mathbf{r}) = \frac{(-i \sqrt{2} \ell_{\kappa_z})^{n_a+n_b}}{\sqrt{2\pi\ell_{\kappa_z}^2n_a!n_b!L}} e^{+ \bar{\mathfrak{z}}_s \mathfrak{z}_s/4\ell_{\kappa_z}^2} \frac{\partial^{n_b}}{\partial \mathfrak{z}_s} \frac{\partial^{n_a}}{\partial \bar{\mathfrak{z}}_s}e^{- \bar{\mathfrak{z}}_s \mathfrak{z}_s/2\ell_{\kappa_z}^2} e^{i(\kappa_z-\beta s[n_a-n_b])z}.
\end{equation}
A technical clarification is in order. The appearance of $e^{-i\beta s[n_a-n_b]z}$ does not originate directly from the algebra above. We have inserted this by hand to ensure that the resulting state remains an eigenstate of $\hat{\kappa}_z$ with the same eigenvalue $\kappa_z$ with which we began. This is necessary because the $\hat{a}$ and $\hat{b}$ operators are only defined \textit{within} a $\kappa_z$ sector. Consequently, it cannot map states outside of that sector. If we insist that the extra exponential factors not be added \textit{ad hoc}, we can modify the definition of the $\hat{a}$ and $\hat{b}$ operators to explicitly account for them. To do this, we note that $\hat{a}^\dagger$ adds $s$ unit of angular momentum. So, it should become, in the real space representation, $\hat{a}^\dagger \mapsto e^{-is \beta z} \hat{a}^\dagger.$ On the other hand, $\hat{b}^\dagger$ depletes $s$ unit of angular momentum. So, it should become, in the real space representation, $\hat{b}^\dagger \mapsto e^{+is \beta z} \hat{b}^\dagger.$ These phases clearly preserve the commutation relations, and do not change the form of the Hamiltonian. The only effect comes when writing down the wavefunctions. These phases ensure that states at with $\kappa_z$ remain inside the $\kappa_z$ sector when acted upon repeatedly by $\hat{a}^\dagger$ and $\hat{b}^\dagger.$ We shall leave these phases off in the primary discussion, multiplying them in only when necessary. In the twisted frame, when working with $k_z$ and not $\kappa_z,$ these phases are never necessary because $k_z$ is completely decoupled from the in-plane degrees of freedom. This technical remark applies equally well to the case of $\kappa_z = 0$. Now, Eq. \eqref{eq: LG wf 1} can be simplified by the use of associated Laguerre polynomials $L_{n_b}^{n_a-n_b}(\chi) = \frac{e^\chi \chi^{-(n_a-n_b)}}{n_b!} \frac{\partial^{n_b}}{\partial \chi} \left( e^{-\chi}\chi^{n_a} \right)$
\begin{equation}
    \begin{split}
        \frac{\partial^{n_b}}{\partial \mathfrak{z}_s} \frac{\partial^{n_a}}{\partial \bar{\mathfrak{z}}_s}e^{- \bar{\mathfrak{z}}_s \mathfrak{z}_s/2\ell_{\kappa_z}^2} &= \frac{(-1)^{n_a}}{\bar{\mathfrak{z}}_s^{n_a}}\frac{\partial^{n_b}}{\partial \mathfrak{z}_s}\left(\frac{\mathfrak{z}_s \bar{\mathfrak{z}}_s}{2\ell^2_{\kappa_z}} \right)^{n_a}e^{- \bar{\mathfrak{z}}_s \mathfrak{z}_s/2\ell_{\kappa_z}^2} = \frac{(-1)^{n_a}}{\bar{\mathfrak{z}}_s^{n_a}}\left( \frac{\bar{\mathfrak{z}}_s}{2\ell_{\kappa_z}^2} \right)^{n_b}\frac{\partial^{n_b}}{\partial \chi } \chi^{n_a}e^{- \chi} \\
        &=\frac{(-1)^{n_a} n_b!}{\bar{\mathfrak{z}}_s^{n_a}}\left( \frac{\bar{\mathfrak{z}}_s}{2\ell_{\kappa_z}^2} \right)^{n_b} e^{- \bar{\mathfrak{z}}_s \mathfrak{z}_s/2\ell_{\kappa_z}^2} \left(\frac{\mathfrak{z}_s \bar{\mathfrak{z}}_s}{2\ell^2_{\kappa_z}} \right)^{n_a-n_b} L_{n_b}^{n_a-n_b}\left( \frac{\mathfrak{z}_s \bar{\mathfrak{z}}_s}{2\ell^2_{\kappa_z}}\right) \\
        &= \frac{(-1)^{n_a} n_b!}{\left(2\ell_{\kappa_z}^2 \right)^{n_a}} e^{- \bar{\mathfrak{z}}_s \mathfrak{z}_s/2\ell_{\kappa_z}^2} \mathfrak{z}_s^{n_a-n_b} L_{n_b}^{n_a-n_b}\left( \frac{\mathfrak{z}_s \bar{\mathfrak{z}}_s}{2\ell^2_{\kappa_z}}\right).
    \end{split}
\end{equation}
We thus find \cite{robinett2006quantum}
\begin{equation}
\label{eq: GL functions}
\begin{split}
    \psi_{n_a,n_b}(\mathbf{r}) &=(-i)^{n_a+n_b} (-1)^{n_a} \sqrt{\frac{n_b!}{n_a!}} \frac{e^{- r^2/4\ell_{\kappa_z}^2}}{\sqrt{2\pi\ell_{\kappa_z}^2L}} \left( \frac{r}{\sqrt{2}\ell_{\kappa_z}}\right)^{n_a-n_b} e^{is \left[n_a-n_b\right]\phi} L_{n_b}^{n_a-n_b}\left( \frac{r^2}{2\ell^2_{\kappa_z}}\right)e^{i(\kappa_z-\beta s[n_a-n_b])z}\\
    &\sim e^{- r^2/4\ell_{\kappa_z}^2} r^{n_a+n_b} e^{is \left[n_a-n_b\right]\phi} e^{i(\kappa_z-\beta s[n_a-n_b])z} =  e^{- r^2/4\ell_{\kappa_z}^2} r^{N} e^{iM\phi} e^{i(\kappa_z-\beta M)z},
\end{split}
\end{equation}
where $N = n_a+n_b$ and $M = s(n_a-n_b).$ The $\kappa_z = 0$ case can be derived analogously. This explicitly shows that the basis states are exponentially localized. Of course, these basis states are \textit{not} eigenstates of the free-particle Hamiltonian. It seems rather formidable  to write eigenstates of the Hamiltonian in this basis as we know they are delocalized Bessel functions. However, the localized basis is convenient, especially for potentially describing crystalline lattices. We therefore would like to make connection between the two. 

To make progress, we must weakly confine the free electrons. Presumably, in a crystal, there is some confining potential present. So, this addition is physically motivated. In fact, in order to normalize the Bessel functions, we need to impose a large box; therefore, in any case, some type of regularization is needed. The simplest confining potential is an isotropic trap:
\begin{equation}
    \hat{\mathcal{V}}_0 = \mathcal{V}_0 \left( \hat{x}^2  + \hat{y}^2\right),
\end{equation}
where we define a confining length scale $q^{-1}:$ $\mathcal{V}_0 = \hbar^2\beta^2 q^2 /2m.$ For definiteness, let us choose $\beta q>0$ (actually, to simplify the algebra, one can also assume that $\beta >0$ and $q>0$ individually). Unlike before, we do \textit{not} need to keep track of the sign of $\beta q.$ This confining potential, being isotropic, has the same form in either reference frame. The Hamiltonian now is 
\begin{equation}
    \begin{split}
        \hat{\mathcal{K}} &= \frac{1}{2m} \left[ \hat{p}^2_{x} + \hat{p}^2_{y} + \hbar^2 \beta^2 q^2  \left( \hat{x}^2 + \hat{y}^2\right)+ \left[\hbar \kappa_z - \beta \left( \hat{x} \hat{p}_{y} - \hat{y} \hat{p}_{x} \right) \right]^2  \right] \\
        &= \frac{1}{2m} \left[ \left(\hat{p}_{x} + \hbar\beta q  \hat{y} \right)^2+ \left(\hat{p}_{y}- \hbar\beta  q \hat{x}\right)^2 +2\hbar\beta q \left( \hat{x} \hat{p}_y   - \hat{y}\hat{p}_x  \right) + \left[\hbar \kappa_z - \beta \left( \hat{x} \hat{p}_{y} - \hat{y} \hat{p}_{x} \right) \right]^2  \right].
    \end{split}
\end{equation}
We can write operators exactly as before
\begin{equation}
\label{eq: equivalence between a b and x y part 2}
    \begin{pmatrix}
        \hat{\mathfrak{a}} \\
        \hat{\mathfrak{a}}^\dagger \\
        \hat{\mathfrak{b}} \\
        \hat{\mathfrak{b}}^\dagger
    \end{pmatrix} = \frac{\ell_{q}}{\sqrt{2}\hbar}\begin{pmatrix}
        1 & -i &-i & -1 \\
         1 & i & i & -1 \\
         1 & -i & i &  1 \\
         1 & i & -i &  1
    \end{pmatrix} \begin{pmatrix}
        \hat{p}_x \\
        \hbar \beta q\hat{x}\\
        \hat{p}_y \\
        \hbar \beta q\hat{y}
    \end{pmatrix} \rightarrow  \begin{pmatrix}
        \hat{p}_x \\
        \hbar \beta q\hat{x}\\
        \hat{p}_y \\
        \hbar \beta q\hat{y}
    \end{pmatrix} = \frac{\hbar}{2\sqrt{2}\ell_{q}}\left(
\begin{array}{cccc}
 1 & 1 & 1 & 1 \\
 i  & -i  & i & -i \\
 i  & -i  & -i & i \\
 -1 & -1 & 1 & 1 \\
\end{array}
\right)\begin{pmatrix}
        \hat{\mathfrak{a}} \\
        \hat{\mathfrak{a}}^\dagger \\
        \hat{\mathfrak{b}} \\
        \hat{\mathfrak{b}}^\dagger
    \end{pmatrix},
\end{equation}
where $\ell_q^2 = \left(2\beta q\right)^{-1}.$ All the preceding considerations go through effortlessly. In particular, we have $\hat{x}\hat{p}_y - \hat{y}\hat{p}_x = \hbar \left( \hat{\mathfrak{a}}^\dagger \hat{\mathfrak{a}} - \hat{\mathfrak{b}}^\dagger \hat{\mathfrak{b}} \right),$ $\frac{1}{2m} \left[ \left(  \hat{p}_x + \hbar\beta q\hat{y} \right)^2 + \left(\hat{p}_y - \hbar\beta q\hat{x} \right)^2 \right] = \hbar \omega_{q} \left(\hat{\mathfrak{b}}^\dagger \hat{\mathfrak{b}} + \frac{1}{2} \right),$ and $\frac{\hbar\beta q }{m} \left( \hat{x} \hat{p}_y   - \hat{y}\hat{p}_x  \right) = \frac{\hbar \omega_q}{2} \left( \mathfrak{a}^\dagger \mathfrak{a} - \mathfrak{b}^\dagger \mathfrak{b} \right).$ Here, $\omega_q = 2\hbar \beta q/m.$ In terms of these new operators, the Hamiltonian is
\begin{equation}
    \hat{\mathcal{K}} = \frac{\hbar \omega_q}{2} \left(  \mathfrak{a}^\dagger \mathfrak{a} + \mathfrak{b}^\dagger \mathfrak{b} + 1 \right) + \frac{\hbar^2 }{2m}\left[ \kappa_z - \beta \left( \hat{\mathfrak{a}}^\dagger \hat{\mathfrak{a}} - \hat{\mathfrak{b}}^\dagger \hat{\mathfrak{b}} \right) \right]^2.
\end{equation}
We can read off the energies immediately
\begin{equation}
\label{eq: energy of confined Hamiltonian}
    E_{N,M}(\kappa_z) = \frac{\hbar^2 \beta q}{m} \left( N +1 \right)  + \frac{\hbar^2 }{2m}\left[ \kappa_z - \beta M \right]^2,
\end{equation}
where $N = n_\mathfrak{a}+n_\mathfrak{b}$ and $M = n_\mathfrak{a}-n_\mathfrak{b}$ are the total mode number and angular momentum defined similar to before. The eigenstates are the same Gauss-Laguerre functions as in Eq. \eqref{eq: GL functions} with $\ell_{\kappa_z}, n_a, n_b, s$ replaced by $\ell_q, n_\mathfrak{a}, n_\mathfrak{b}, +1 $ respectively. We note that this approach differs from the Bogoliubov transformation approach used in the main text in that no special treatment is required for the $\kappa_z = 0$ case. However, the two are equivalent via a gauge transformation.

\subsection{Energy Spectrum}

\begin{figure}
    \centering
    \includegraphics[width=1\linewidth]{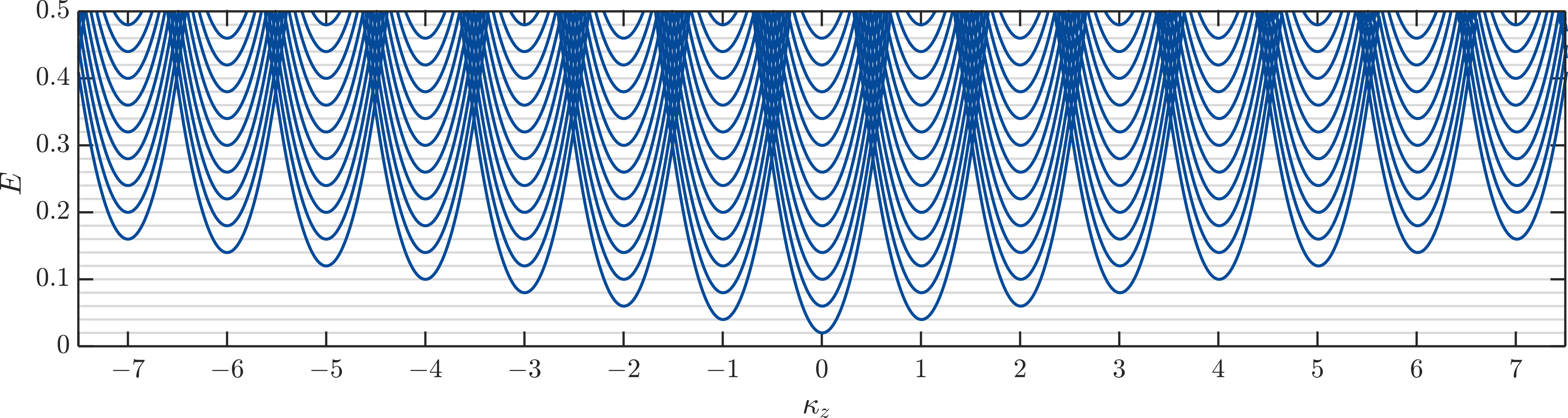}
    \caption{\textbf{Spectrum of  free particles weakly confined in an isotropic trap.}  The lowest-energy state occurs at $E = 2q$ at $\kappa_z = 0.$ This state has both $N = M = 0.$  The vertex of each quadratic branch corresponds to its angular momentum. For instance, a branch with vertex $\kappa_z = 2$ has angular momentum $M = 2.$ Branches with different $N$ values but have the same angular momentum $M$ are shifted upwards relative to each other in energy by increments of $4q$ (i.e. differing by $\pm 2$ in $N$). The light gray horizontal lines are placed at increments of $2q$ in energy. Here, $q = 0.01.$}
    \label{fig:free spectrum}
\end{figure}

We end this section by studying the energy spectrum. For convenience, we scale momenta by $\beta$ and energies by $\hbar^2\beta^2/2m$ (assuming that $\beta >0$).  We find the renormalized energy is
\begin{equation}
\label{eq: dimensionless spectrum}
    E_{N,M}(\kappa_z) = 2 q \left(N + 1 \right) +   \left( \kappa_z - M\right)^2.
\end{equation}
To avoid too many variables, we continue to use $\kappa_z,$ $q$ and $E$ to mean both dimensionless and dimensionful quantities. It should be clear from context which definition is employed in a particular instance. Recalling that $N = n_\mathfrak{a}+ n_\mathfrak{b}$ and $M = n_\mathfrak{a}- n_\mathfrak{b},$ where $n_\mathfrak{a}$ and $n_\mathfrak{b}$ can be any non-negative integer, we find that $(N\pm  M)/2 \geq 0$ must be an integer. This implies that $N \geq |M|,$ i.e. for a given $M,$ $N$ must be at least as big as $|M|.$ Furthermore, $M$ and $N$ must be of the same parity since the sum and difference have to be divisible by $2.$ Therefore, $N = |M| + 2n,$ where $n = 0,1,2,3,...$ For a fixed $N,$ the range of $M$ is bounded: $|M| \leq N. $ The bounds are saturated when we have only Bogoliubov quasiparticles of one type. Some representative spectra are shown in Fig. \ref{fig:free spectrum}. From there, we first observe that if all values of $N$ are included, the spectra are dense in energy. This must be the case since we are plotting a four-dimensional data set projected onto a two-dimensional surface. In other words, because the energy is characterized by three quantum numbers, we know that the spectrum is dense when plotted against just one of them (in this case, $\kappa_z$). For a fixed $N,$ the minimum energy $E_\text{min} = 2q(N+1).$ For a fixed $N$ and $M,$ the band dispersion is quadratic in $\kappa_z$ with vertex centered at $\kappa_z = M.$ For a fixed $M,$ the spectrum at different values of $N > |M|$ are just simply shifted upward in energy by $4q.$ If we are interested in an energy window $\left[0,W\right],$ then roughly only modes with $N < W/q$ contribute to the  physics. It makes sense that the weaker the confinement potential, the more modes must be included in the energy window.

\section{Effect of Saddle-Point Potential Energy}
\label{sec: Effect of Saddle-Point Potential Energy}

\begin{figure}
    \centering
    \includegraphics[width=0.5\linewidth]{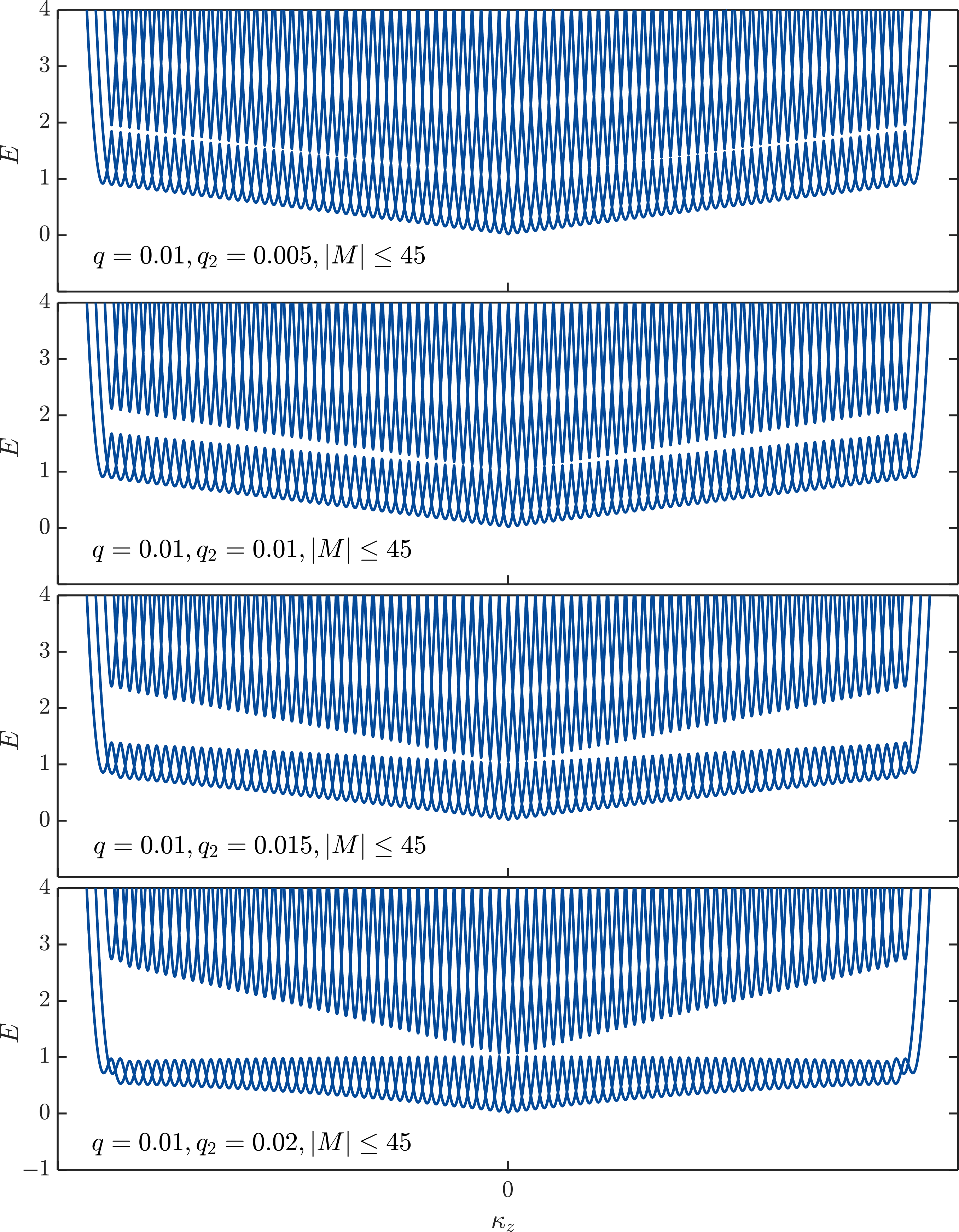}
    \caption{\textbf{Spectrum in the presence of saddle-point potential energy with only minimal $N$ included.} In these calculations, only states with $N = |M| \leq 45$ are included. As $q_2$ increases, four "edge" modes become better and better developed. }
    \label{fig:bandstructure1}
\end{figure}

\begin{figure}
    \centering
    \includegraphics[width=0.5\linewidth]{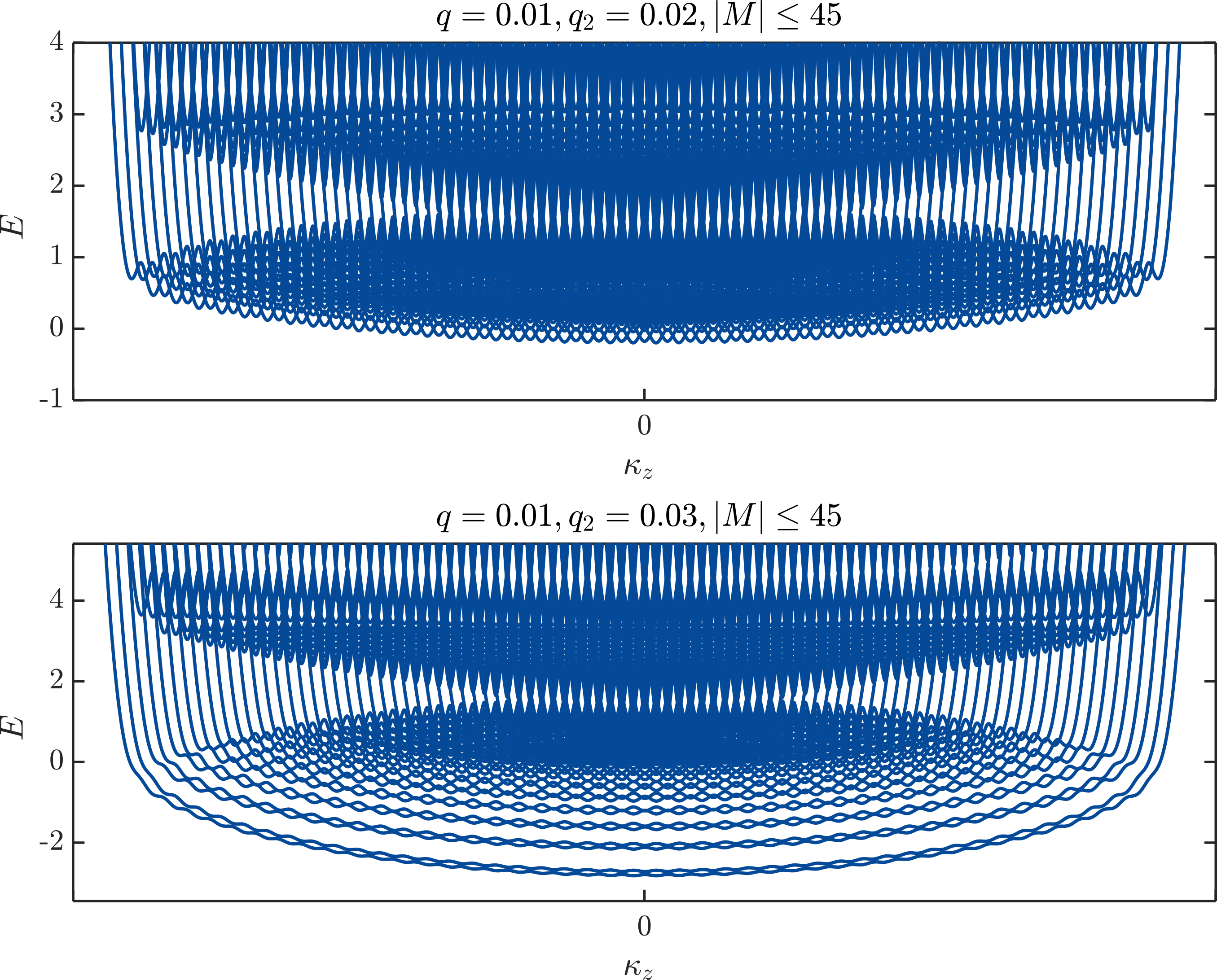}
    \caption{\textbf{Spectrum in the presence of saddle-point potential energy.} In these calculations, only states with $N  \leq 45$ are included. In other words, at a fixed $M,$ we include all $|M|\leq N \leq 45.$  }
    \label{fig:bandstructure2}
\end{figure}

We now consider the effect of a saddle-point potential energy on the energy spectrum of the weakly-confined particles. The potential energy has the following representation in the twisted frame
\begin{equation}
    \mathcal{V}_2  = -\frac{\hbar^2 \beta^2 q_2^2}{2m} \left( x_\text{twist}^2- y_\text{twist}^2 \right) = -\frac{\hbar^2 \beta^2 q_2^2}{2m} \frac{1}{2}\left[ \left( x_\text{twist}+iy_\text{twist}\right)^2 +\left( x_\text{twist}-iy_\text{twist}\right)^2 \right].
\end{equation}
In the laboratory frame, this same potential energy has a more complicated form 
\begin{equation}
    \mathcal{V}_2  = -\frac{\hbar^2 \beta^2 q_2^2}{2m} \frac{1}{2}\left[ (x_\text{lab}-iy_\text{lab})^2 e^{2i\beta z_\text{lab}} + (x_\text{lab}+iy_\text{lab})^2 e^{-2i\beta z_\text{lab}} \right] = -\frac{\hbar^2 \beta^2 q_2^2}{2m} \frac{r^2}{2}\left[  e^{2i\beta z_\text{lab}-2i \phi_\text{lab}} +   e^{2i \phi_\text{lab}-2i\beta z_\text{lab}} \right].
\end{equation}
It is straightforward check that in the twisted frame, $\hat{\mathcal{V}}_2$ commutes with $\hat{p}_{z_\text{twist}}$ since it is independent of $z_\text{twist}.$ In the laboratory frame, $\hat{\mathcal{V}}_2$ commutes with $\hat{\kappa}_z$ (again, this shows that $\hat{p}_{z_\text{twist}}$ and $\hbar\hat{\kappa}_z$ are the same operator; so if $\hat{\mathcal{V}}$ commutes with it in one frame, it must commute in any other frame). So we can write $\mathcal{V}_2$ in each $\kappa_z$ or $k_z$ sector. We can now rewrite $\hat{\mathcal{V}}_2$ in terms of the ladder operators. First, let us do that for the $\hat{a}$ and $\hat{b}$ operators. Recalling that $\hat{x} = \frac{i \ell_{\kappa_z}}{\sqrt{2}} \left(\hat{a} - \hat{a}^\dagger + \hat{b} - \hat{b}^\dagger \right) $ and $\hat{y} = \frac{s \ell_{\kappa_z}}{\sqrt{2}} \left(-\hat{a} - \hat{a}^\dagger + \hat{b} + \hat{b}^\dagger \right),$ we find $\hat{x} + i s\hat{y} = i \sqrt{2}\ell_{\kappa_z} \left(\hat{b}-\hat{a}^\dagger \right).$ Consequently, we find, in either frame,
\begin{equation}
    \hat{\mathcal{V}}_2 = \frac{\hbar^2 \beta^2 q_2^2}{2m} \ell^2_{\kappa_z} \left[ \left( \hat{b} - \hat{a}^\dagger\right)^2  + \left( \hat{b}^\dagger - \hat{a}\right)^2 \right].
\end{equation}
In terms of the $\hat{\mathfrak{a}}$ and $\hat{\mathfrak{b}}$  operators or the $\hat{\mathbb{a}}$ and $\hat{\mathbb{b}}$ operators, the potential energy is  the same with only one replacement $\ell_{\kappa_z} \mapsto \ell_{q}$
\begin{equation}
    \hat{\mathcal{V}}_2 = \frac{\hbar^2 \beta^2 q_2^2}{2m} \ell^2_{q} \left[ \left( \hat{\mathfrak{b}} - \hat{\mathfrak{a}}^\dagger\right)^2  + \left( \hat{\mathfrak{b}}^\dagger - \hat{\mathfrak{a}}\right)^2 \right] = \frac{\hbar^2 \beta^2 q_2^2}{2m} \ell^2_{q} \left[ \left( \hat{\mathbb{b}} - \hat{\mathbb{a}}^\dagger\right)^2  + \left( \hat{\mathbb{b}}^\dagger - \hat{\mathbb{a}}\right)^2 \right].
\end{equation}
Because the $\hat{\mathfrak{a}}$ and $\hat{\mathfrak{b}}$ operators do not carry cumbersome factors of $s,$ we choose to write the matrix elements of $\hat{\mathcal{V}}_2$ in terms of these operators. First, we note the usual actions of the operators on the basis states:
\begin{equation}
    \begin{split}
        \hat{\mathfrak{a}}^\dagger \ket{n_\mathfrak{a}, n_\mathfrak{b}} &= \sqrt{n_\mathfrak{a}+1} \ket{n_\mathfrak{a}+1, n_\mathfrak{b}}, \\
        \hat{\mathfrak{b}}^\dagger \ket{n_\mathfrak{a}, n_\mathfrak{b}} &= \sqrt{n_\mathfrak{b}+1} \ket{n_\mathfrak{a}, n_\mathfrak{b}+1}, \\
        \hat{\mathfrak{a}} \ket{n_\mathfrak{a}, n_\mathfrak{b}} &= \sqrt{n_\mathfrak{a}} \ket{n_\mathfrak{a}-1, n_\mathfrak{b}}, \\
        \hat{\mathfrak{b}} \ket{n_\mathfrak{a}, n_\mathfrak{b}} &= \sqrt{n_\mathfrak{b}} \ket{n_\mathfrak{a}, n_\mathfrak{b}-1},
    \end{split}
\end{equation}
Then, writing $\ket{N,M} = \ket{n_\mathfrak{a} = \frac{N+M}{2}, n_\mathfrak{b} = \frac{N-M}{2}},$ we have
\begin{equation}
    \begin{split}
        \hat{\mathfrak{a}}^\dagger \ket{N, M} &= \sqrt{\frac{N+M}{2}+1} \ket{N+1, M+1}, \\
        \hat{\mathfrak{b}}^\dagger \ket{N, M}  &= \sqrt{\frac{N-M}{2}+1} \ket{N+1, M-1}, \\
        \hat{\mathfrak{a}} \ket{N, M}  &= \sqrt{\frac{N+M}{2}} \ket{N-1, M-1}, \\
        \hat{\mathfrak{b}} \ket{N, M}  &= \sqrt{\frac{N-M}{2}} \ket{N-1, M+1}.
    \end{split}
\end{equation}
Evaluating actions on  the quadratic terms, we find 
\begin{equation}
    \begin{split}
        \left( \hat{\mathfrak{b}} - \hat{\mathfrak{a}}^\dagger\right)^2 \ket{N,M} = &\left(  \left[\hat{\mathfrak{b}}\right]^2 - 2\hat{\mathfrak{b}}\hat{\mathfrak{a}}^\dagger +   \left[\hat{\mathfrak{a}}^\dagger\right]^2 \right)\ket{N,M} \\ 
        =&\sqrt{\frac{N-M}{2}\left(\frac{N-M}{2}-1 \right)} \ket{N-2,M+2} - 2 \sqrt{\frac{N-M}{2}\left(\frac{N+M}{2}+1 \right)}\ket{N,M+2} + \\
        &+\sqrt{\left(\frac{N+M}{2}+1\right)\left(\frac{N+M}{2}+2 \right)} \ket{N+2,M+2}, \\
        \left( \hat{\mathfrak{b}}^\dagger - \hat{\mathfrak{a}}\right)^2 \ket{N,M} = &\left(  \left[\hat{\mathfrak{b}}^\dagger\right]^2 - 2\hat{\mathfrak{b}}^\dagger\hat{\mathfrak{a}} +   \left[\hat{\mathfrak{a}}\right]^2 \right)\ket{N,M} \\ 
        =&\sqrt{\left(\frac{N-M}{2}+1\right)\left(\frac{N-M}{2}+2 \right)} \ket{N+2,M-2} - 2 \sqrt{\frac{N+M}{2}\left(\frac{N-M}{2}+1 \right)}\ket{N,M-2} + \\
        &+\sqrt{\left(\frac{N+M}{2}\right)\left(\frac{N+M}{2}-1 \right)} \ket{N-2,M-2}.
    \end{split}
\end{equation}
We observe by explicit calculation that this potential mixes angular momenta which differ by $\pm 2$ units. So the spectrum partitions into a mirror-even sector and a mirror-odd sector. Explicitly, the non-zero matrix elements in dimensionless units are
\begin{equation}
    \begin{split}
        \bra{N-2,M+2} \hat{\mathcal{V}_2} \ket{N,M} &= \frac{q_2^2}{2q}\sqrt{\frac{N-M}{2}\left(\frac{N-M}{2}-1 \right)}, \\
        \bra{N,M+2} \hat{\mathcal{V}_2} \ket{N,M} &= -2\frac{q_2^2}{2q}\sqrt{\frac{N-M}{2}\left(\frac{N+M}{2}+1 \right)}, \\
        \bra{N+2,M+2} \hat{\mathcal{V}_2} \ket{N,M} &= \frac{q_2^2}{2q}\sqrt{\left(\frac{N+M}{2}+1\right)\left(\frac{N+M}{2}+2 \right)}, \\
        \bra{N-2,M-2} \hat{\mathcal{V}_2} \ket{N,M} &= \frac{q_2^2}{2q}\sqrt{\left(\frac{N+M}{2}\right)\left(\frac{N+M}{2}-1 \right)}, \\
        \bra{N,M-2} \hat{\mathcal{V}_2} \ket{N,M} &= -2\frac{q_2^2}{2q}\sqrt{\frac{N+M}{2}\left(\frac{N-M}{2}+1 \right)}, \\
        \bra{N+2,M-2} \hat{\mathcal{V}_2} \ket{N,M} &= \frac{q_2^2}{2q}\sqrt{\left(\frac{N-M}{2}+1\right)\left(\frac{N-M}{2}+2 \right)}. \\
    \end{split}
\end{equation}
Using these matrix elements, we calculate the energy spectra of particles in the saddle-point potential energy, weakly regularized by the isotropic confining potential. We only include $|M|$ up to a certain maximum value $M_\text{max}$. Physically, this means that there are ``hard walls" which prohibit the existence of states beyond a certain radius. In the first calculation, we only include $N = |M|.$ In this case, we find, as shown in Fig. \ref{fig:bandstructure1}, the development of four ``edge" states becomes clear as the strength of the saddle-point potential energy increases. If, we include all $N$ values up to $M_\text{max},$ then the energy spectrum is shown in Fig. \ref{fig:bandstructure2}. There, we also see the ``edge" states, but they only become clear for large values of $q_2.$

\section{Boundary modes of Hofstadter networks}
\label{sec: Boundary modes of Hofstadter networks}

In the main text, we discussed the bulk spectra of the Hofstadter networks. These networks consist of a periodic array of saddle points described by an intra-saddle-point tunneling amplitude $t$. Tunneling across neighboring saddle points proceeds the inter-saddle-point tunneling amplitude $t'$. This is illustrated in Fig.\ 4 of the main text.

\begin{figure}
    \centering
    \includegraphics[width=1\linewidth]{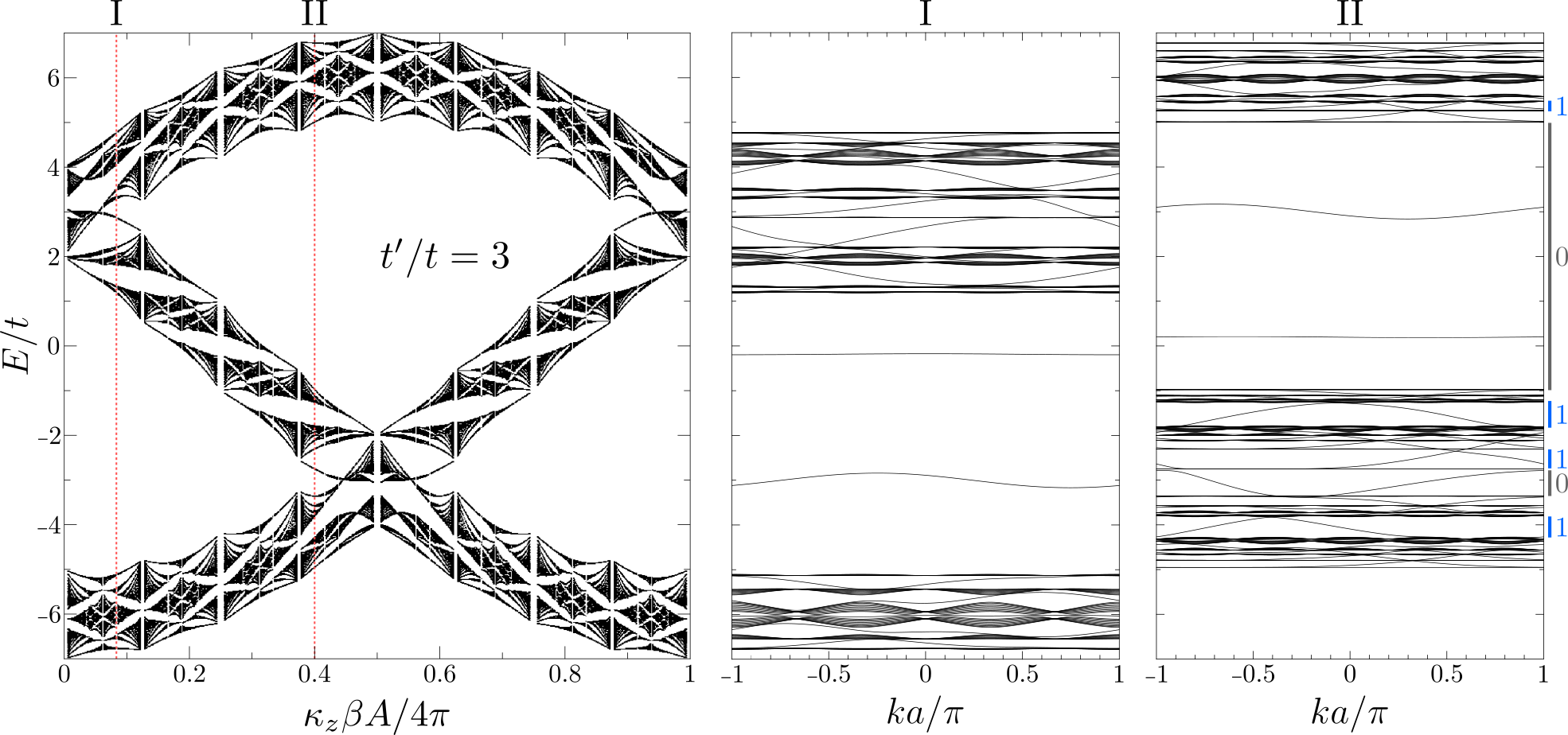}
    \caption{\textbf{Spectrum of a triangular Hofstadter network for strong inter-saddle-point tunneling.} (left) Bulk spectrum as a function of the effective flux $2 \hbar \kappa_z \beta A / e$ in units of the flux quantum where $A$ is the unit cell area. (right) Spectrum for ribbon of width given by $10$ magnetic unit cells for the indicated values of the effective flux. Some Chern numbers (for a fixed $\kappa_z$ sector) are indicated on the rightmost panel, corresponding to the net number of chiral modes at a given edge.}
    \label{fig:network1}
\end{figure}

\begin{figure}
    \centering
    \includegraphics[width=1\linewidth]{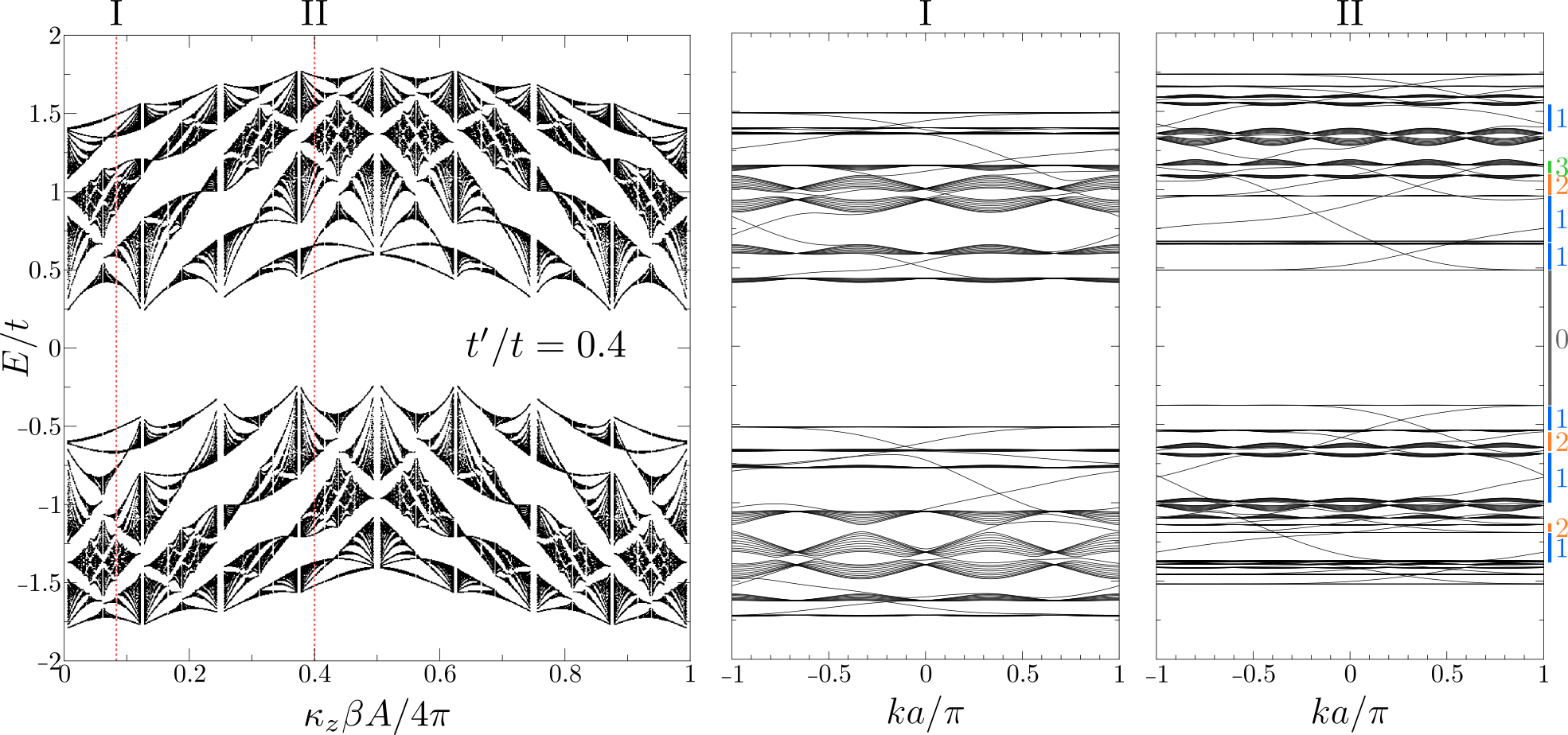}
    \caption{\textbf{Spectrum of a triangular Hofstadter network for weak inter-saddle-point tunneling.} (left) Bulk spectrum as a function of the effective flux $2 \hbar \kappa_z \beta A / e$ in units of the flux quantum where $A$ is the unit cell area. (right) Spectrum for ribbon of width given by $10$ magnetic unit cells for the indicated values of the effective flux. Some Chern numbers (for a fixed $\kappa_z$ sector) are indicated on the rightmost panel, corresponding to the net number of chiral modes at a given edge.}
    \label{fig:network2}
\end{figure}

In Figs.\ \ref{fig:network1} and \ref{fig:network2}, we show the spectrum for the bulk and finite-width ribbons of the triangular Hofstadter network, to illustrate the boundary modes appearing in the bulk Hofstadter gaps. We consider both the limit of strong (Fig.\ \ref{fig:network1}) and weak (Fig.\ \ref{fig:network2}) inter-saddle-point tunneling using the same parameter values as in the main text.
\textbf{}

\twocolumngrid

\bibliography{SM_References}

\end{document}